\documentclass[10pt,conference]{IEEEtran}
\IEEEoverridecommandlockouts
\usepackage[table]{xcolor}
\usepackage{cite}
\usepackage{multirow}
\usepackage{graphicx}
\usepackage{booktabs}
\usepackage{makecell}
\usepackage{pifont}
\usepackage[framemethod=TikZ]{mdframed}
\usepackage{pgf}
\usepackage{collcell}
\usepackage{algorithm} 
\usepackage{algorithmic}

\newcommand{\ApplyGradientA}[1]{%
    \pgfmathsetmacro{\PercentColor}{100.0*#1/40}
    \edef\x{\noexpand\cellcolor{orange!\PercentColor}}\x{#1}%
}
\newcommand{\ApplyGradientB}[1]{%
    \pgfmathsetmacro{\PercentColor}{100.0*#1/40}
    \edef\x{\noexpand\cellcolor{orange!\PercentColor}}\x{#1}%
}
\newcommand{\ApplyGradientC}[1]{%
    \pgfmathsetmacro{\PercentColor}{100.0*#1/50}
    \edef\x{\noexpand\cellcolor{orange!\PercentColor}}\x{#1}%
}
\newcommand{\ApplyGradientD}[1]{%
    \pgfmathsetmacro{\PercentColor}{100.0*#1/3.1}
    \edef\x{\noexpand\cellcolor{orange!\PercentColor}}\x{#1}%
}

\newcolumntype{A}{>{\collectcell\ApplyGradientA}{c}<{\endcollectcell}}
\newcolumntype{B}{>{\collectcell\ApplyGradientB}{c}<{\endcollectcell}}
\newcolumntype{C}{>{\collectcell\ApplyGradientC}{c}<{\endcollectcell}}
\newcolumntype{D}{>{\collectcell\ApplyGradientD}{c}<{\endcollectcell}}

\def\BibTeX{{\rm B\kern-.05em{\sc i\kern-.025em b}\kern-.08em
    T\kern-.1667em\lower.7ex\hbox{E}\kern-.125emX}}
\begin{document}

\title{An Empirical Study on Commit Message Generation using LLMs via In-Context Learning}

\author{
\IEEEauthorblockN{
Yifan Wu\IEEEauthorrefmark{1},
Yunpeng Wang\IEEEauthorrefmark{2},
Ying Li\IEEEauthorrefmark{1},
Wei Tao\IEEEauthorrefmark{3},
Siyu Yu\IEEEauthorrefmark{1},
Haowen Yang\IEEEauthorrefmark{4},
Wei Jiang\IEEEauthorrefmark{2},
Jianguo Li\IEEEauthorrefmark{2}
}
\IEEEauthorblockA{\IEEEauthorrefmark{1}Peking University, Beijing, China}
\IEEEauthorblockA{\IEEEauthorrefmark{2}Ant Group, Hangzhou, China}
\IEEEauthorblockA{\IEEEauthorrefmark{3}Fudan University, Shanghai, China}
\IEEEauthorblockA{\IEEEauthorrefmark{4}The Chinese University of Hong Kong, Shenzhen (CUHK-Shenzhen), China}
\thanks{This work was done when Yifan Wu was an intern at Ant Group. Ying Li is the corresponding author.}}

\maketitle

\begin{abstract}
Commit messages concisely describe code changes in natural language and are important for software maintenance. Several approaches have been proposed to automatically generate commit messages, but they still suffer from critical limitations, such as time-consuming training and poor generalization ability.
To tackle these limitations, we propose to borrow the weapon of large language models (LLMs) and in-context learning (ICL). Our intuition is based on the fact that the training corpora of LLMs contain extensive code changes and their pairwise commit messages, which makes LLMs capture the knowledge about commits, while ICL can exploit the knowledge hidden in the LLMs and enable them to perform downstream tasks without model tuning. 
However, it remains unclear how well LLMs perform on commit message generation via ICL.
In this paper, we conduct an empirical study to investigate the capability of LLMs to generate commit messages via ICL.
Specifically, we first explore the impact of different settings on the performance of ICL-based commit message generation.
We then compare ICL-based commit message generation with state-of-the-art approaches on a popular multilingual dataset and a new dataset we created to mitigate potential data leakage.
The results show that ICL-based commit message generation significantly outperforms state-of-the-art approaches on subjective evaluation and achieves better generalization ability. 
We further analyze the root causes for LLM's underperformance and propose several implications, which shed light on future research directions for using LLMs to generate commit messages.
\end{abstract}


\section{Introduction}
When submitting a code change to a version control system, developers can write a brief descriptive comment in the format of natural language, called a commit message. 
High-quality commit messages can greatly facilitate software maintenance by providing a human-readable summary of the changes and the rationale behind them, obviating the need for a detailed examination of the complex code and potentially simplifying code review and related tasks \cite{mockus2000identifying,tao2012software}. Conversely, poor commit messages can negatively affect software defect proneness \cite{li2023commit}.

However, the inherent complexity of commits renders the task of manually summarizing them into concise messages difficult and prone to errors \cite{dyer2013boa,maalej2010can}. According to Tian et al. \cite{tian2022makes}, 44\% of the commit messages from five open-source projects have quality issues. 
Furthermore, within the context of today's accelerated software development pace, manually writing high-quality commit messages becomes a time-intensive and arduous task \cite{tao2021evaluation}. Research indicates that a significant amount of commit messages from open-source projects lack important information \cite{tian2022makes} or are even empty \cite{dyer2013boa}.

Consequently, many approaches have been proposed to generate high-quality commit messages automatically over the years. 
Early studies \cite{buse2010automatically, cortes2014automatically, linares2015changescribe, shen2016automatic} extract information from code changes and generate commit messages with predefined rules or templates, which may not encompass all scenarios or deduce the intent behind code changes. 
Later, some studies \cite{huang2017mining, liu2018neural, huang2020learning} adopt information retrieval (IR) techniques to reuse commit messages of similar code changes. They can take advantage of similar examples, but the reused commit messages may not correctly describe the content or intent of the current code change. 
Recently, a large number of learning-based approaches have been proposed \cite{liu2020atom, nie2021coregen, wang2021context, dong2022fira, jung2021commitbert, shi2022race, he2023come,tao2024kadel, lin2023cct5}. They trained on a large-scale commit corpus to translate code changes into commit messages and have been demonstrated to outperform rule-based approaches and IR-based approaches.

Although learning-based approaches have achieved comparative success, they still have some critical limitations. 
First, these approaches require either training models from scratch \cite{liu2020atom, nie2021coregen, wang2021context, dong2022fira} or tuning a pre-trained model \cite{jung2021commitbert, shi2022race, he2023come, tao2024kadel, lin2023cct5} (e.g., CodeT5 \cite{wang2021codet5}) with labeled data, which could be impractical due to the scarcity of computing resources and labeled data.
Second, the performance of these approaches largely depends on the quality of training data, while the most widely-used training data, i.e., commits from open-source projects, is typically of poor quality \cite{tian2022makes, li2023commit}.
Third, these approaches suffer from significant performance degradation when applied to new projects due to poor generalization ability. Research indicates that the performance of learning-based approaches can decrease by 26.93\% to 73.41\% in such a scenario \cite{tao2022large}.

To address the above limitations, we propose to borrow the weapon of large language models (LLMs) and in-context learning (ICL).
LLMs are pre-trained on extensive unlabeled corpora via self-supervised learning, thereby acquiring substantial knowledge. 
Considering the inclusion of abundant real-world commits within the corpora, LLMs present a promising avenue for commit message generation.
In addition, ICL, i.e., providing a prompt with demonstrations to LLMs, has been shown to effectively harness the knowledge inherent in LLMs and enable them to perform downstream tasks without model tuning \cite{dong2022survey}.
Some recent works \cite{eliseeva2023commit, tao2024kadel, zhang2024using, lopes2024commit, zhang2024automatic, wu2024commit} have investigated LLM-based commit message generation. However, it remains unclear how well LLMs perform on commit message generation via ICL. More research is needed to determine its ability in this important area.

Therefore, in this paper, we conduct an empirical study on commit message generation using LLMs via ICL.
Specifically, considering the sensitivity of LLMs to different settings, we first investigate the impact of different prompts and demonstrations on the performance of ICL-based commit message generation.
Based on the optimal setting, we compare various LLMs with state-of-the-art baselines on a popular benchmark \cite{tao2022large}.
To mitigate the potential data leakage, we create a new dataset by collecting commits from repositories not included in the benchmark (named MCMD-NL) and recent commits from the same repositories included in the benchmark (named MCMD-NT), respectively. 
We evaluate LLMs and baselines on the above datasets using both objective metrics and subjective metrics.
Based on the results, we perform an in-depth analysis of the root causes of LLM's underperforming cases.

Our study highlights the capability of LLMs to generate commit messages through ICL and identifies several directions for future research. The key findings are as follows:
1) Prompt settings have a greater impact on ICL-based commit message generation in zero-shot learning than in few-shot learning, suggesting that demonstrations can mitigate the LLM’s sensitivity to prompt variations. 
2) A moderate number of demonstrations can enhance the performance of ICL-based commit message generation, but an excessive number can reduce performance.
3) Retrieval-based demonstration selection can statistically significantly enhance the performance of ICL-based commit message generation, while the order of demonstrations has minimal impact on performance.
4) GPT-3.5-Turbo and DeepSeek-V2-Chat are the best-performing LLMs for the commit message generation task, statistically outperforming other LLMs on all metrics.
5) The best-performing LLMs statistically significantly outperform the best-performing baseline on MCMD-NT, indicating superior generalization. Moreover, the best-performing LLMs have comparable performance on MCMD-NL to the best-performing baseline, which is fine-tuned on the training set.
6) In the subjective evaluation, the best-performing LLMs statistically significantly outperform the best-performing baseline. Additionally, LLM-based evaluation has much higher correlations with human judgment than objective metrics, suggesting it is more reliable to evaluate the quality of commit messages.
7) LLMs still struggle in some cases, mainly due to a lack of contextual knowledge, adverse demonstrations, and model fallacy. Providing high-quality demonstrations and more advanced LLMs can resolve most underperforming cases.

In summary, this paper makes the following contributions:
\begin{itemize}
    \item We conduct an empirical study to explore the performance of ICL-based commit message generation.
    \item We analyze the root causes of underperforming cases and propose important directions for future research.
    \item The code and dataset in this study are publicly available at https://github.com/wuyifan18/LLM4CMG to benefit both practitioners and researchers in the field of commit message generation.
    
\end{itemize}

\section{Background}

\subsection{Commit, Diff, and Commit Messages}
\begin{figure}
    \centering
    \includegraphics[width=0.95\columnwidth]{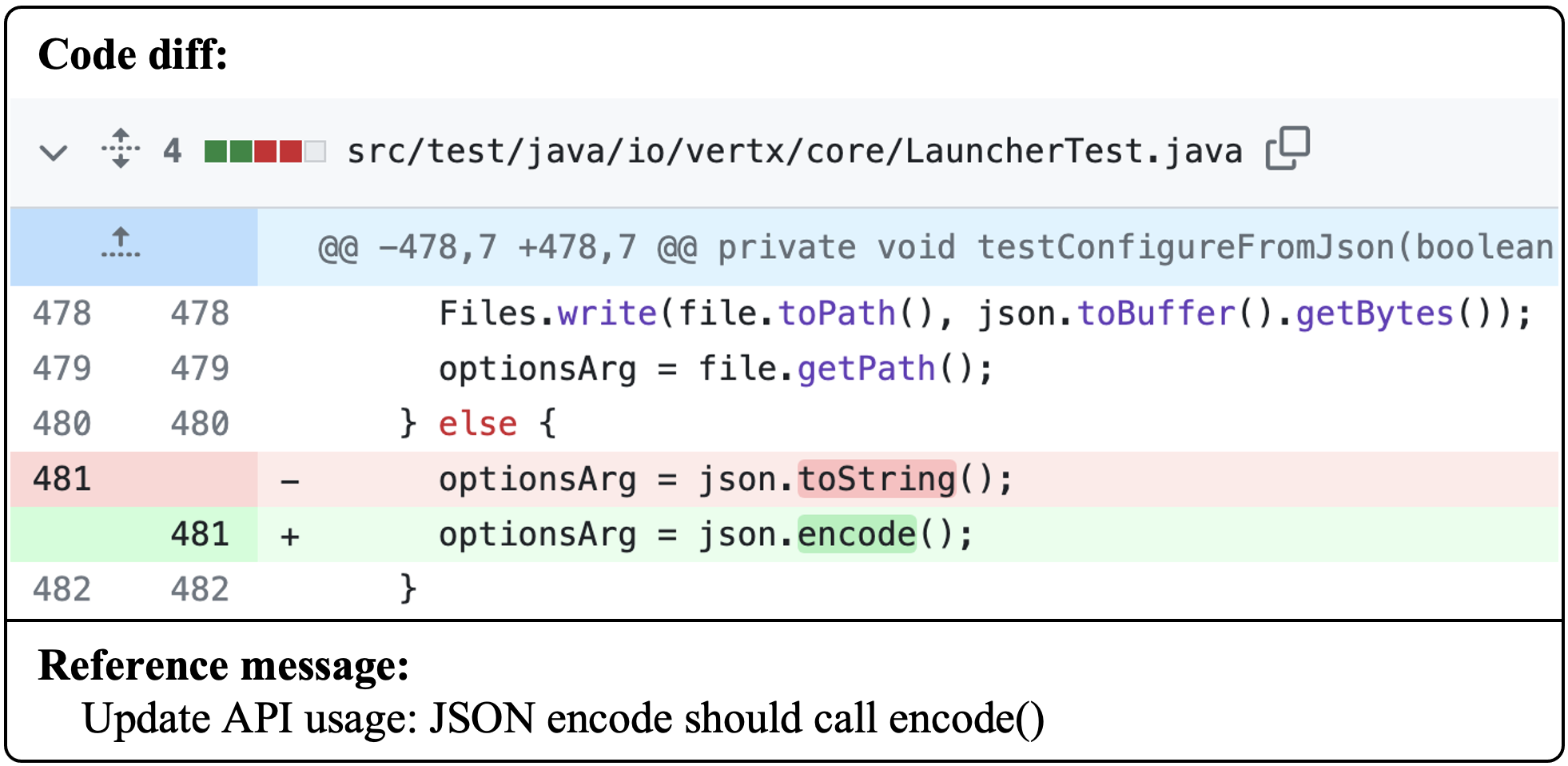}
    \caption{An example of a commit with a code \emph{diff} and its reference message.}
    \label{fig:example}
    \vspace{-2ex}
\end{figure}

Git \cite{git} is one of the most popular version control systems. Whenever developers submit a code change, Git will create a commit to record this change and require developers to enter a commit message.
The commit message is written by developers in a textual format to facilitate the understanding of the change, while the change here is represented by \emph{diff}, which characterizes the difference between two versions. 
Usually, a diff contains one or multiple chunks with file paths. The modified codes are wrapped by “@@” in a chunk with the negative sign “-” or positive sign “+” with a line number to denote the deleted or added lines of code.
As shown in \figurename~\ref{fig:example}, a commit mentioned in this paper refers to the pair of a code \emph{diff} and its corresponding commit message. Given a commit, we refer to its original commit message as \emph{reference message}. 

\begin{figure*}
    \centering
    \includegraphics[width=0.95\textwidth]{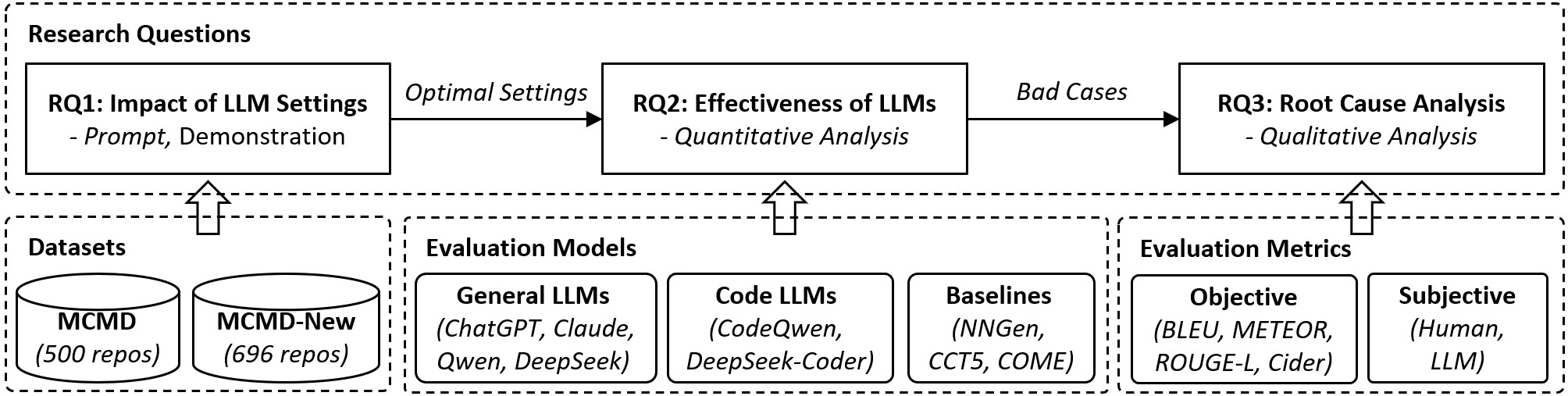}
    \caption{Overview of our study.}
    \label{fig:overview}
    \vspace{-2ex}
\end{figure*}

\subsection{Large Language Models}
Large Language Models (LLMs) are large-sized pre-trained language models with tens or hundreds of billions of parameters trained on extensive unlabeled corpora via self-supervised learning, which exhibit strong capacities to understand natural language and solve various tasks \cite{zhao2023survey}.
In addition to natural language, LLMs can also deal with code, which arouses growing interest in applying LLMs to the software engineering domain \cite{hou2023large, fan2023large}.
How to effectively apply LLMs to downstream tasks has become an important research topic. A prevalent paradigm is to fine-tune the model by learning the desired output from the given input of a downstream dataset to update model parameters \cite{wang2021codet5}.
For example, to enhance the performance and adaptability of LLMs in specific tasks within the software engineering domain, there are a considerable amount of fine-tuned code LLMs (e.g., Codex \cite{chen2021evaluating} and Code Llama \cite{roziere2023code}), and have made tremendous achievements in various tasks, such as commit message generation \cite{he2023come}, code review \cite{li2022automating}, and just-in-time comment update \cite{lin2023cct5}.

\subsection{In-Context Learning}
Tuning a pre-trained model on downstream datasets can be extremely time-consuming and resource-intensive, especially for LLMs that usually contain billions of parameters. In addition, the performance of tuned models largely depends on the scale and quality of labeled data, making it less feasible in specific scenarios with limited labeled data.
Recently, in-context learning (ICL) offers a new paradigm that allows LLMs to perform downstream tasks without model tuning \cite{dong2022survey}, where a formatted natural language prompt is used as the input for LLMs to generate the output.
Specifically, the prompt typically includes three parts: (1) Instruction: the description of the specific task; (2) Demonstrations: a few examples (i.e., query-answer pairs) selected from the task datasets; (3) Query: the test query that LLMs need to answer.
Such a prompt can let LLMs gain task-specific knowledge by learning the pattern hidden in the demonstrations of the task.
Many studies have demonstrated that LLMs achieved remarkable performance in various tasks via ICL, such as log statement generation \cite{xu2024unilog} and comment generation \cite{geng2024large}.

\section{Study Design}

\subsection{Overview and Research Questions}
\figurename~\ref{fig:overview} shows the overview of our study.
We evaluate popular LLMs and state-of-the-art commit message generation approaches on a popular multilingual benchmark dataset MCMD \cite{tao2022large}. 
However, given the potential risk that the dataset could be used to train LLMs and baselines, we create a new dataset MCMD-New consisting of two parts: new commits from the same repositories as MCMD but collected more recently and commits from repositories using different languages that are not included MCMD.
We next introduce the research questions we aim to investigate and their relationships.

\begin{itemize}
    \item \textbf{RQ1: Impact of LLM Settings: How do different prompt and demonstration settings affect the performance of ICL-based commit message generation?} 
    Recent studies have shown that the performance of ICL is highly affected by prompt and demonstration settings \cite{dong2022survey, liu2023pre}. Inspired by these studies, we propose to investigate the impact of LLM settings.
    Specifically, we design four prompts based on whether a role description is provided and whether constraint information is given and investigate demonstration settings from three dimensions (i.e., number, selection, and order).
    \item \textbf{RQ2: Effectiveness of LLM: How does the performance of ICL-based commit message generation compare with state-of-the-art approaches?}
    Based on the optimal setting obtained from RQ1, in this RQ, we evaluate the performance of ICL-based commit message generation compared with state-of-the-art approaches.
    To answer this question, we compare various LLMs with state-of-the-art approaches using both objective metrics and subjective metrics on the existing MCMD dataset and the new dataset MCMD-New we created.
    \item \textbf{RQ3: Root Cause Analysis: What are the underlying causes for LLM's underperformance?}
    In this RQ, we aim to further investigate the root causes of LLM’s underperforming cases.
    Specifically, we sample a diverse set of 200 underperforming cases where LLM failed to make accurate predictions in RQ2 and summarize the categories of root causes. 
\end{itemize}

\begin{figure}
    \centering
    \includegraphics[width=0.95\columnwidth]{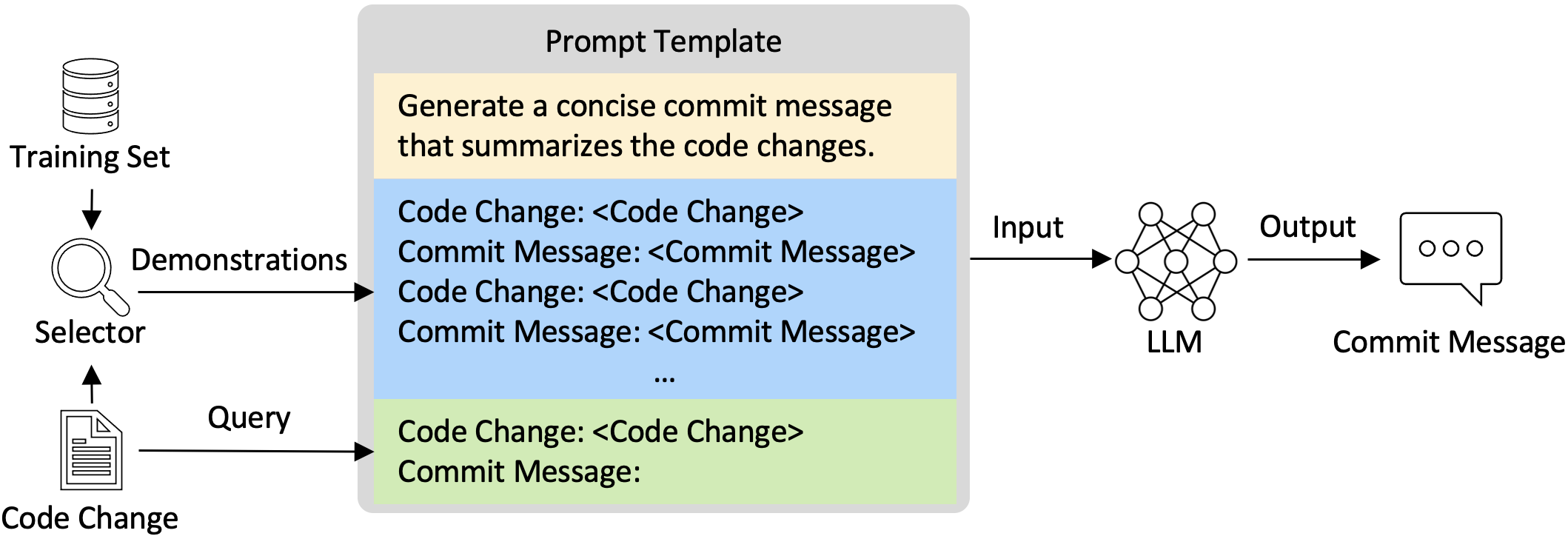}
    \caption{Overview of ICL-based commit message generation.}
    \label{fig:framework}
    \vspace{-2ex}
\end{figure}

\subsection{Prompt Templates for Commit Message Generation}
Formally, a prompt is defined as $\mathcal{P}=\{\mathcal{NL}+\mathcal{CD}+x_q\}$, where $\mathcal{NL}$ is a natural language instruction,  $\mathcal{CD}=\{(x_i,y_i)\}^n_{i=1}$ is a set of code change demonstrations composed by input code change sequence $x_i$ and desired output sequence $y_i$, and $x_q$ is a query to be answered by LLMs. 
Specifically, if $n=0$ which means there is no demonstration, the setting is known as \emph{zero-shot learning}; if $n=1$ which means there is only one demonstration, the setting is known as \emph{one-shot learning}; and \emph{few-shot learning} means there are several demonstrations. Also, there is a constraint that the prompt should fit within the context window limit of LLMs.

\figurename~\ref{fig:framework} illustrates a structured prompt template for commit message generation. The template begins with a natural language instruction $\mathcal{NL}$, as highlighted in the \emph{yellow} text. 
The instruction is followed by a set of code change demonstrations $\mathcal{CD}$, highlighted in the \emph{blue} text. The demonstrations consist of one code change and one associated commit message.
By analogizing the demonstrations, LLM can learn the expected behavior and directly generate a commit message for the given query.
Finally, the template concludes with a code change query $x_q$ for LLM to generate the desired commit message, illustrated at the bottom of the template.

\subsection{Demonstration Selection}
\label{sec:selection}
Given a code change query $x_q$, we select a set of demonstrations $\mathcal{CD}$, assemble them into a prompt $\mathcal{P}$, and input it into LLM for inference. The whole process is shown in \figurename~\ref{fig:framework}.
We next introduce the demonstration selection methods we used.

\subsubsection{Random-based Selection}
This method randomly selects a set of code change demonstrations from the training set, which is the most straightforward and efficient method. 

\subsubsection{Retrieval-based Selection}
\label{sec:retrieval}
Many studies have shown that demonstrations that are similar to the query may help LLMs better understand the desired behavior \cite{nashid2023retrieval, gao2023makes, geng2024large}.
To this end, we use a \emph{k}NN-based demonstration selection algorithm that does not involve much computational overhead in practice, which is shown in Algo. \ref{alg}.
For a code change query $x_q$, we calculate its similarity $sim(x_q,x_i)$ with each code change candidate $x_i$ from the training set. Then, we select top-\emph{k} most similar code changes as demonstrations.
We utilize three methods to calculate similarity as follows:

\begin{itemize}
    \item \textbf{Token-based:} The most widely-used method to retrieve similar code is focusing on the overlap of tokens \cite{golubev2021multi}. Inspired by these studies, we utilize the Jaccard Coefficient \cite{niwattanakul2013using} to calculate the similarity at token level as follows: $sim(x_q,x_i)=\frac{| F(x_q) \cap F(x_i)|}{|F(x_q) \cup F(x_i)|}$, where $F(\cdot)$ calculates the number of tokens in a code change.
    \item \textbf{Frequency-based:} BM-25 \cite{robertson2009probabilistic}, which is an extension of TF-IDF \cite{ramos2003using}, is a classic sparse retrieval method in the information retrieval field and also used in code intelligence tasks \cite{nashid2023retrieval}. Therefore, our second method utilizes BM-25 to calculate the similarity. We implement BM-25 with the gensim package \cite{vrehuuvrek2010software}.
    \item \textbf{Semantic-based:} The above two methods can only employ the lexical similarity, recent studies have revealed that the code semantic is also important to find similar code \cite{zeng2023degraphcs}.
    Hence, we use OpenAI embedding model \cite{embedding} to embed the query $x_q$ and candidate $x_i$ into vector representations $v_q$ and $v_i$. The similarity is then quantified as the cosine similarity, i.e., $sim(x_q,x_i)=\frac{v_q \cdot v_i}{\|v_q\|_2 \|v_i\|_2}$.
\end{itemize}

\begin{algorithm}[t]
\caption{\emph{k}NN-based Demonstration Selection} 
\label{alg}  
\begin{algorithmic}[1] 
    \REQUIRE query $x_q$, training set $\mathcal{D}={(x_i,y_i)}^N_{i=1}$, similarity function ${sim(\cdot)}$, demonstration number \emph{k}
    \ENSURE Demonstrations $\mathcal{CD}$
    \STATE $\mathcal{C}=\emptyset$, $\mathcal{CD}=\emptyset$
    \FOR{$x_i \in \mathcal{D}$}
        \STATE $s_{i} = sim(x_q,x_i)$
        \STATE $\mathcal{C} = \mathcal{C} \cup \{i:s_i\}$
    \ENDFOR
    \STATE extract keys with top-\emph{k} largest values from $\mathcal{C}$ to $\mathcal{I}$
    \FOR{$i \in \mathcal{I}$}
        \STATE $\mathcal{CD}=\mathcal{CD} \cup (x_i, y_i)$
    \ENDFOR
    \STATE \textbf{return} $\mathcal{CD}$
\end{algorithmic} 
\end{algorithm}

\begin{table}
\centering
\caption{The statistics of MCMD-New dataset.}
\label{tab:dataset}
\begin{tabular}{clcc}
\toprule
Dataset         & Language & Repo. & Commit \\ \midrule
\multirow{5}{*}{MCMD-NT} & C++               & 68             & 83,887          \\
                         & C\#               & 74             & 43,046          \\
                         & Java              & 63             & 34,704          \\
                         & Python            & 82             & 36,427          \\
                         & JavaScript        & 80             & 31,428          \\ \midrule
\multirow{5}{*}{MCMD-NL} & PHP               & 76             & 31,395          \\
                         & R                 & 51             & 9,624           \\
                         & TypeScript        & 88             & 83,765          \\
                         & Swift             & 75             & 8,295           \\
                         & Objective-C       & 39             & 2,620           \\ 
\bottomrule
\end{tabular}
\vspace{-2ex}
\end{table}
\subsection{Datasets}

\subsubsection{MCMD}
We first select the MCMD dataset\cite{tao2022large}, which is widely used in the commit message generation task. This dataset contains five programming languages: C++, C\#, Java, Python, and JavaScript. For each language, it collects commits from the top 100 most-starred repositories on GitHub. Shi et al. \cite{shi2022race} further filter out commits with files that cannot be parsed (such as .jar, .ddl, .mp3, and .apk) to reduce noise data and build a higher-quality dataset, containing 500 repositories and 1,094,115 commits. We use the dataset in our experiment.

\subsubsection{MCMD-New}
Due to the potential overlap with the training data of LLMs resulting in data leakage, we created a new dataset MCMD-New.
Specifically, we collected data after January 1, 2022, as the training data of LLMs (e.g., ChatGPT) is up until September 2021 \cite{ChatGPT}. Furthermore, MCMD does not contain data after January 1, 2022.
In addition to the repositories included in MCMD, we crawled commits from additional 500 repositories (top 100 most-starred repositories for each language) using five programming languages: PHP, R, TypeScript, Swift, and Objective-C, which are not included in MCMD and enjoy widespread popularity according to the PYPL PopularitY index \cite{PYPL}.
In total, we selected 1,000 repositories, with 500 repositories from MCMD and 500 new repositories with different programming languages.

To mitigate the presence of low-quality commits which could affect the comparisons between LLMs and baselines,
we employed regular regressions to filter low-quality commit messages, including redundant messages (e.g., rollback commits) and noisy messages (e.g., update files), which is consistent with that used in previous studies \cite{liu2018neural, shi2022race}.
After applying the filtering rules and selecting commits based on time, we get 696 repositories out of the initial 1,000 repositories.
As shown in Table~\ref{tab:dataset}, MCMD-New comprises two parts: MCMD-NewTime (MCMD-NT), which includes 229,492 commits from 367 repositories also present in MCMD, and MCMD-NewLanguage (MCMD-NL), which includes 135,699 commits from 329 new repositories that have different programming languages from the repositories in MCMD.
Finally, following the data splitting of prior studies \cite{tao2022large, he2023come}, we randomly divide MCMD-New into training set, validation set, and test set, with proportions of 80\%, 10\%, and 10\%, respectively.

\subsection{Evaluation Models}

\subsubsection{LLMs}
We select six LLM families with widespread popularity and exceptional performance in code generation \cite{liu2024your}. The summary of studied LLMs is shown in Table \ref{tab:LLM_summary}.

\textbf{ChatGPT \cite{ChatGPT}.}
ChatGPT is the most representative LLM released by OpenAI and has garnered significant attention in the field of software engineering \cite{hou2023large, fan2023large}. We employ GPT-3.5-Turbo in our experiments.

\textbf{Claude 3 \cite{anthropic2024claude}.}
Claude 3 is a family of large multimodal models developed by Anthropic. We utilize Claude-3-Haiku, the fastest and most affordable model in our experiments.

\textbf{Qwen1.5 \cite{qwen1.5}.}
Qwen1.5 released by Alibaba Cloud is a family of language models with different model sizes. We utilize CodeQwen1.5-7B-Chat in our experiments.

\textbf{DeepSeek-V2 \cite{deepseekv2}.}
DeepSeek-V2 is a family of Mixture-of-Experts (MoE) language models, which is pre-trained on a corpus comprising 8.1 trillion tokens. We utilize DeepSeek-V2-Chat in our experiments.

\textbf{CodeQwen1.5 \cite{CodeQwen}.}
CodeQwen1.5 is a family of code LLMs, which is built upon Qwen1.5 and further pre-trained with around 3 trillion tokens of code-related data. We utilize CodeQwen1.5-7B-Chat in our experiments.

\textbf{DeepSeek-Coder-V2 \cite{zhu2024deepseek}.}
DeepSeek-Coder-V2 is a family of MoE code LLMs, which is further pre-trained based on DeepSeek-V2 with an additional 6 trillion tokens. We utilize DeepSeek-Coder-V2-Instruct in our experiments.

\subsubsection{Baselines}
We select the following state-of-the-art commit message generation approaches as baselines to compare with LLMs.

\textbf{NNGen \cite{liu2018neural}.} NNGen is a state-of-the-art retrieval-based commit message generation approach \cite{tao2022large}. It represents code change as bag-of-words vectors \cite{manning2008introduction}. The vector of the test code change is then compared to those in the training set using cosine similarity. The commit message of the most similar code change in the training set is reused as the result.

\textbf{CCT5 \cite{lin2023cct5}.} CCT5 is the state-of-the-art code-change-oriented pre-trained model, which is built on top of T5 model \cite{raffel2020exploring}. It is pre-trained with 39.6GB of code change data collected from 35K GitHub repositories. Five code-change-oriented tasks are used for pre-training. We fine-tune and evaluate the pre-trained model on our training and test sets.

\textbf{COME \cite{he2023come}.} COME is another state-of-the-art commit message generation approach. It uses modification embedding to represent code changes and fine-tuned CodeT5 \cite{wang2021codet5} to generate a translation result and retrieve a result simultaneously. The final commit message is selected from the above two results with an SVM-based decision algorithm.

\begin{table}
\centering
\caption{The summary of LLMs used in our study.}
\label{tab:LLM_summary}
\resizebox{\columnwidth}{!}{%
\begin{tabular}{@{}ccccc@{}}
\toprule
Category & Family & Model & Open-Source & Release Date \\ \midrule
\multirow{6}{*}{General LLMs} & ChatGPT & GPT-3.5-Turbo & \ding{55} & 06/2023 \\ \cmidrule(l){2-5} 
 & Claude 3 & Claude-3-Haiku & \ding{55} & 03/2024 \\ \cmidrule(l){2-5} 
 & Qwen1.5 & Qwen1.5-7B-Chat & \ding{51} & 02/2024 \\ \cmidrule(l){2-5} 
 & DeepSeek-V2 & DeepSeek-V2-Chat & \ding{51} & 05/2024 \\ \midrule
\multirow{2}{*}{Code LLMs} & CodeQwen1.5 & CodeQwen1.5-7B-Chat & \ding{51} & 04/2024 \\ \cmidrule(l){2-5} 
 & DeepSeek-Coder-V2 & DeepSeek-Coder-V2-Instruct & \ding{51} & 06/2024 \\ \bottomrule
\end{tabular}%
}
\vspace{-2ex}
\end{table}

\subsection{Evaluation Metrics}
\label{sec:metrics}

\subsubsection{Objective Evaluation Metrics}
Objective evaluation metrics assess the quality of the generated message by calculating the text similarity between the generated message and the reference message.
We use four widely-used metrics in previous literature \cite{liu2020atom, wang2021context, tao2022large, shi2022race, he2023come}, including BLEU \cite{papineni2002bleu}, METEOR \cite{banerjee2005meteor}, ROUGE-L \cite{lin2004rouge}, and Cider \cite{vedantam2015cider}.
BLEU measures the precision of n-grams between the generated text and the reference texts. 
ROUGE-L is a recall-oriented metric that measures the longest common subsequence between the generated text and the reference texts. 
METEOR calculates the harmonic mean of 1-gram precision and recall of the generated text against the reference texts.
Cider takes each sentence as a document and calculates the cosine similarity of its TF-IDF vector at the n-gram level to obtain the similarity between the generated text and the reference texts. 

\subsubsection{Subjective Evaluation Metrics} 
Objective evaluation metrics fail to capture the semantics of the generated messages and have been shown to have a relatively low correlation with human judgments \cite{tao2022large, liu2023g}. Furthermore, these metrics treat reference messages as the gold standard, which can be of poor quality. To address these issues, we turn to reference-free evaluation methods.

\textbf{Human Evaluation.}
Following the previous work \cite{tao2024kadel, shi2022race, he2023come}, we invite three volunteers, including postgraduates majoring in computer science and industry professionals with relevant experience.
For each sample, we provide volunteers with the code change, the reference message, and messages generated by different models. To ensure unbiased evaluation, we shuffled the commit messages so that the volunteers did not know where each message came from.
Each participant is asked to assign scores from 1 to 5 (the higher the better) to the commit messages from three aspects: Informativeness (the extent of important information about the code change reflected in the commit message), Conciseness (the extent of extraneous information included in the commit message), and Expressiveness (readability and fluency of the commit message). The final score for one commit message is the average of the three volunteers.

\begin{figure}
    \centering
    \includegraphics[width=0.95\columnwidth]{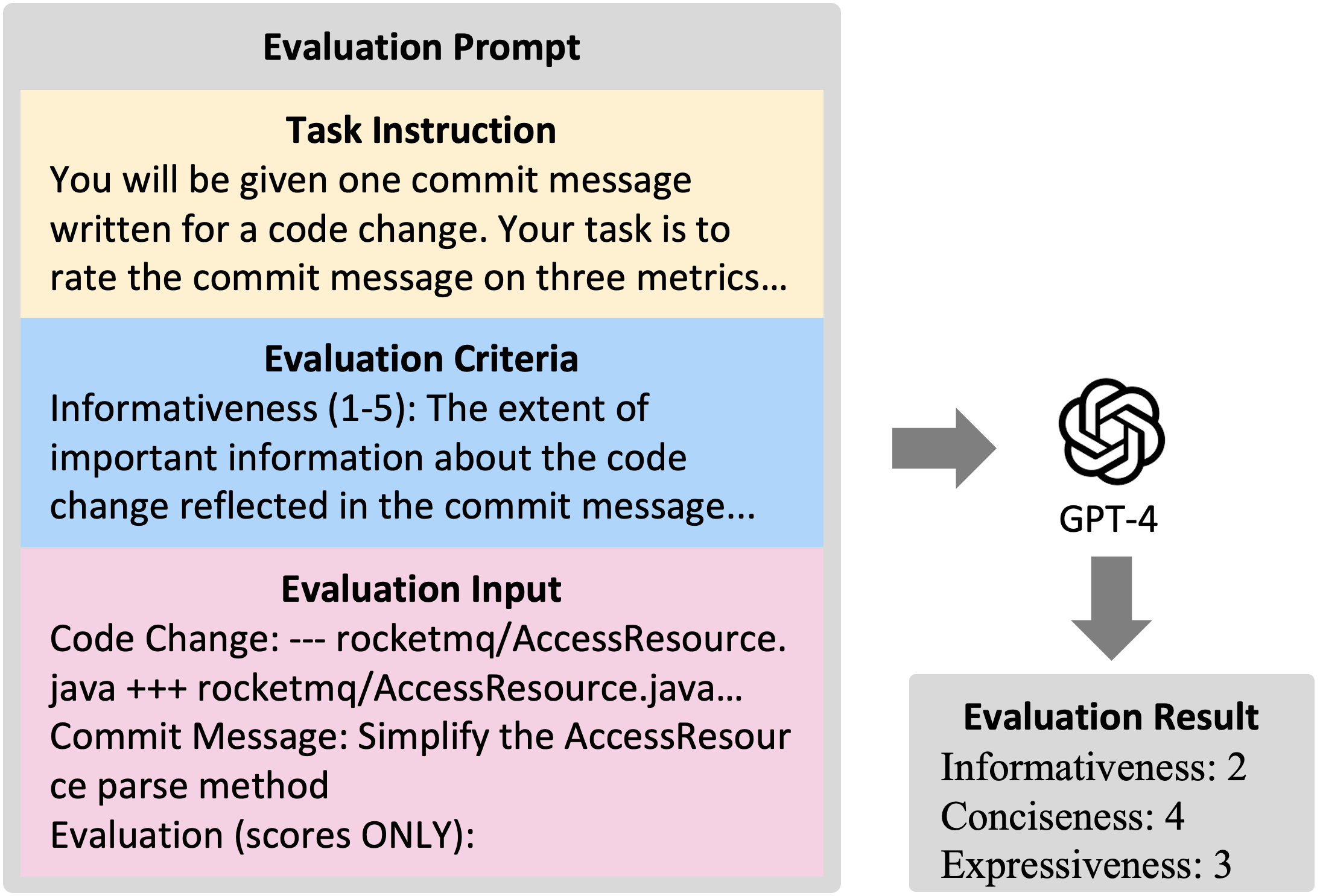}
    \caption{The prompt for LLM-based evaluation.}
    \label{fig:g-eval}
    \vspace{-2ex}
\end{figure}

\textbf{LLM-based Evaluation.}
In addition to human evaluation, we utilize LLM-based evaluation to evaluate the quality of generated commit messages.
Recent studies indicate that LLM as an evaluator achieves high alignment with human \cite{liu2023g, zheng2024judging}.
To this end, we use GPT-4 \cite{OpenAI23GPT4} as the evaluator and design an evaluation prompt as the input of GPT-4 to score the generated commit message. As shown in \figurename~\ref{fig:g-eval}, the prompt comprises three parts: 1) a task instruction that contains the definition of the evaluation task; 2) evaluation criteria that describe the detailed standards which is the same as human; 3) an evaluation input that needs to be scored.

\begin{figure*}
    \centering
    \includegraphics[width=0.95\textwidth]{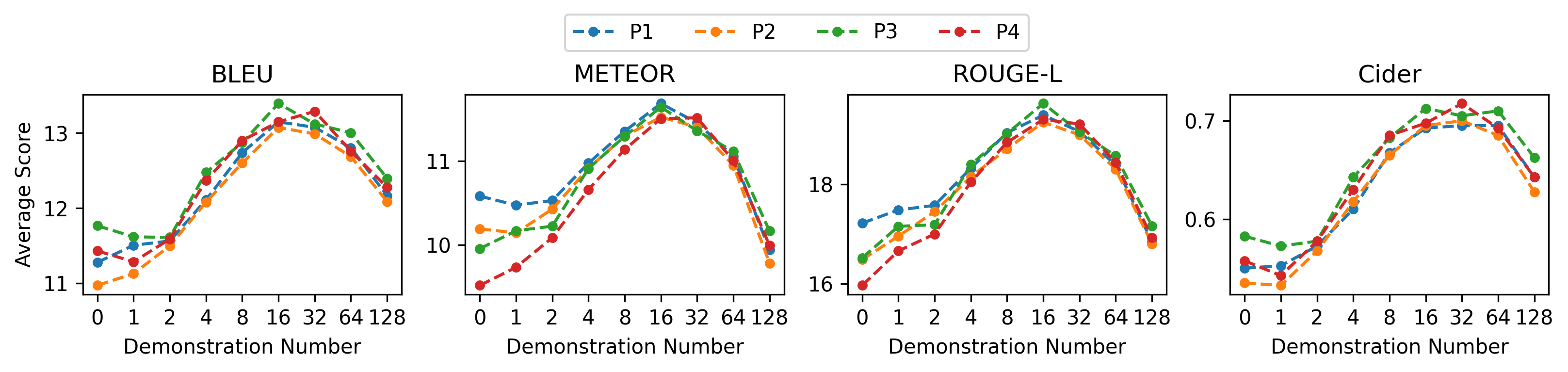}
    \caption{Impact of different prompts and demonstration number on the performance of ChatGPT.}
    \label{fig:prompt}
    \vspace{-2ex}
\end{figure*}

\subsection{Experiment Settings}
\label{impl}
For open-source models such as CodeQwen1.5-7B-Chat, we use official releases available on HuggingFace \cite{huggingface}. For closed-source models (e.g., GPT-3.5-Trubo), we access them via official API. Our experiments are conducted on a Linux server equipped with 4 NVIDIA V100 GPUs. For open-source models, we deploy a local API server using vLLM \cite{kwon2023efficient}, which is a unified library for LLM serving and inference. All models are used in their original precisions without quantization.

Temperature controls the randomness in the generated results of models \cite{chen2021evaluating}. Following previous work \cite{nashid2023retrieval, gao2023makes}, we set the temperature to 0 to ensure deterministic output.
The context length of models limits the input size. Hence, we truncate each demonstration to $\frac{context\_length}{N+1}$ tokens, where $N$ represents the number of demonstrations.
To obtain reliable results, experiments for all open-source models are repeated five times, and we report the average value as the final result. Due to budget constraints, We do not repeat experiments for closed-source models in RQ2.

\section{Study Results}

\subsection{RQ1: Impact of Prompts and Demonstrations}

\subsubsection{Setup}
To select the prompts, we followed the best practices \cite{prompts}, which suggests that prompts consist of four elements: Instruction, Context, Input Data, and Output Indicator. We have tried prompts with various combinations of these four elements. During our preliminary evaluation, we experimented with a total of 10 prompts. Due to budget constraints, we selected the 4 best-performing and representative prompts:
\begin{itemize}
    \item \textbf{Prompt 1 (P1): the simplest prompt.} We only provided the basic requirement of generating a commit message based on the code change without additional description.
    \item \textbf{Prompt 2 (P2): P1 + Role Description.} P2 was designed based on P1 but included a role description that asked LLM to act as a commit message generator and generate a commit message for a code change.
    \item \textbf{Prompt 3 (P3): P1 + Constraint Information.} P3 included constraint information, such as not writing explanations and just replying with the commit message.
    \item \textbf{Prompt 4 (P4): P3 + Role Description.} P4 was a combination of P2 and P3, containing both role description and constraint information.
\end{itemize}

To explore the impact of demonstrations, we first vary the demonstration number from 0 to 128 using random-based selection. We evaluated the impact of 36 various combinations of these 4 prompts and 9 kinds of demonstration numbers.
With the optimal prompt and demonstration number, we further investigated the impact of demonstration selection using the selection methods described in Sec. \ref{sec:retrieval}.
Furthermore, some studies \cite{kumar2021reordering, zhao2021calibrate, lu2022fantastically} suggest that the demonstration order can affect the performance of ICL.
Therefore, we arrange demonstrations in both ascending and descending order based on similarity to explore the impact of demonstration order.

We use ChatGPT and DeepSeek-V2 as the representative LLMs in this RQ. Due to the prohibitive cost of API access, we randomly selected 200 samples from the valid set of the MCMD dataset to reduce the number of API calls. To mitigate the randomness of LLM inference, we repeated each setting five times and reported their mean as the final result.

\begin{table*}
\centering
\caption{Impact of demonstration selection and order on the performance of ChatGPT.}
\label{tab:demonstration}
\resizebox{\textwidth}{!}{%
\begin{tabular}{cccccccccccccccccccccccccc}
\toprule
\multicolumn{2}{c}{Demonstration} & \multicolumn{4}{c}{Java} & \multicolumn{4}{c}{C\#} & \multicolumn{4}{c}{C++} & \multicolumn{4}{c}{Python} & \multicolumn{4}{c}{JavaScript} & \multicolumn{4}{c}{Average} \\ \cmidrule(r){1-2} \cmidrule(r){3-6} \cmidrule(r){7-10} \cmidrule(r){11-14} \cmidrule(r){15-18} \cmidrule(r){19-22} \cmidrule(r){23-26}  
Selection & Order & BLEU & Met. & Rou. & Cid. & BLEU & Met. & Rou. & Cid. & BLEU & Met. & Rou. & Cid. & BLEU & Met. & Rou. & Cid. & BLEU & Met. & Rou. & Cid. & BLEU & Met. & Rou. & Cid. \\ \midrule
Random & Random & 13.80 & 11.66 & 20.53 & 0.80 & 12.45 & 10.45 & 18.23 & 0.63 & 11.47 & 10.40 & 16.88 & 0.54 & 13.81 & 12.39 & 19.88 & 0.67 & 15.44 & 13.30 & 22.62 & 0.94 & 13.40 & 11.64 & 19.63 & 0.71 \\ \midrule
\multirow{2}{*}{Token} & Descend & 22.32 & 19.84 & 29.23 & \textbf{1.59} & 18.30 & 16.20 & 25.16 & 1.15 & 16.33 & 14.68 & 23.02 & \textbf{0.96} & \textbf{17.49} & \textbf{16.21} & \textbf{25.19} & \textbf{1.02} & 21.89 & 18.52 & 29.75 & 1.57 & \textbf{19.26} & 17.09 & 26.47 & 1.25 \\
 & Ascend & \textbf{22.44} & \textbf{20.03} & \textbf{29.85} & 1.58 & 18.59 & 16.28 & 25.09 & 1.22 & 15.85 & 14.39 & 22.37 & 0.89 & 17.14 & 15.78 & 24.98 & 1.00 & \textbf{22.26} & \textbf{19.24} & \textbf{30.74} & \textbf{1.58} & \textbf{19.26} & \textbf{17.14} & \textbf{26.60} & \textbf{1.26} \\ \midrule
\multirow{2}{*}{Frequency} & Descend & 21.86 & 19.74 & 29.55 & 1.52 & 17.36 & 15.05 & 23.80 & 1.13 & 14.93 & 13.55 & 21.39 & 0.84 & 15.98 & 14.26 & 23.65 & 0.89 & 19.95 & 16.99 & 27.34 & 1.38 & 18.02 & 15.92 & 25.15 & 1.15 \\
 & Ascend & 20.29 & 18.09 & 27.21 & 1.43 & 18.43 & \textbf{16.33} & 25.10 & \textbf{1.23} & 16.07 & 14.47 & 22.72 & 0.92 & 15.47 & 14.23 & 22.89 & 0.90 & 20.87 & 17.75 & 28.17 & 1.43 & 18.23 & 16.17 & 25.22 & 1.18 \\ \midrule
\multirow{2}{*}{Semantic} & Descend & 21.45 & 19.13 & 28.33 & 1.54 & 17.78 & 15.36 & 24.22 & 1.09 & 16.46 & \textbf{14.94} & 23.13 & 0.95 & 16.67 & 15.37 & 24.58 & 0.96 & 21.08 & 18.06 & 29.30 & 1.48 & 18.69 & 16.57 & 25.91 & 1.20 \\
 & Ascend & 22.14 & 19.84 & 28.68 & 1.58 & \textbf{18.64} & 16.25 & \textbf{25.58} & 1.16 & \textbf{16.67} & 14.66 & \textbf{23.65} & 0.95 & 16.24 & 15.12 & 23.91 & 0.89 & 20.72 & 17.81 & 28.05 & 1.47 & 18.88 & 16.73 & 25.97 & 1.21 \\ \bottomrule
\end{tabular}%
}
\vspace{-2ex}
\end{table*}

\subsubsection{Results} \figurename~\ref{fig:prompt} illustrates the performance of ChatGPT under different prompts and demonstration numbers. Table \ref{tab:demonstration} displays the impact of demonstration selection and order on the performance of ChatGPT.

\textbf{Impact of Prompt.}
We observe that P3 marginally outperforms the other prompts. By default, we adopted P3 in subsequent experiments.
Moreover, different prompts have a greater impact in zero-shot learning compared to few-shot learning, and the impact gradually decreases as the number of demonstrations increases. Specifically, the difference between the best and worst prompt without any demonstrations on ROUGE-L is 1.26 while that with 16 demonstrations is 0.38. This suggests that incorporating demonstrations can help mitigate LLM's sensitivity to prompts.
One possible reason is that demonstrations in few-shot learning provide additional context and examples, which help guide the LLM's responses and reduce its reliance on the specific wording or structure of the prompt. In contrast, in zero-shot learning, the LLM has to understand the task solely based on the prompt, making it more susceptible to prompt variations.

\begin{mdframed}[nobreak=true, skipabove=5pt, skipbelow=5pt]
\textbf{Finding 1:}
Prompt settings have a greater impact on ICL-based commit message generation in zero-shot learning than in few-shot learning, suggesting that demonstrations can mitigate the LLM's sensitivity to prompt variations.
\end{mdframed}

\textbf{Impact of Demonstration Number.}
We observe that the performance of ChatGPT on all metrics increases with the number of demonstrations and achieves a peak at 16. 
For example, the average improvements of 16 demonstrations over no demonstrations are 16.1\%, 15.1\%, 17.2\%, and 26.1\% on BLEU, METEOR, ROUGE-L, and Cider, respectively.
However, the performance suffers from a significant drop when further increasing the number to 128. 
One possible reason is that too many demonstrations can introduce noise and redundancy, as not all demonstrations are equally relevant or useful. This can confuse the LLM and negatively impact its performance.
By default, we adopted 16 demonstrations.

\begin{mdframed}[nobreak=true, skipabove=5pt, skipbelow=5pt]
\textbf{Finding 2:}
A moderate number of demonstrations enhances ICL-based commit message generation performance, but an excessive number can reduce performance.
\end{mdframed}

\textbf{Impact of Demonstration Selection and Order.}
We observe that retrieval-based selection outperforms random-based selection by an average of 46.1\%. 
The Wilcoxon signed-rank test \cite{wilcoxon1992individual} shows that the improvement achieved by retrieval-based selection is statistically significant at the confidence level of 95\%.
Specifically, BLEU improves by 39.7\%, METEOR improves by 42.6\%, ROUGE-L improves by 31.9\%, and Cider improves by 70.2\% on average.
Among the retrieval-based selection, the token-based method performs slightly better than others, but the difference is not statistically significant. Given the computational efficiency, we adopted the token-based method as the default demonstration selection in subsequent experiments.
Additionally, the performance difference between sorting demonstrations in ascending order and descending order is not statistically significant. By default, we sorted demonstrations in ascending order of similarity.

\begin{mdframed}[nobreak=true, skipabove=5pt, skipbelow=5pt]
\textbf{Finding 3:}
Retrieval-based demonstration selection can statistically significantly improve the performance of ICL-based commit message generation, while the order of demonstrations has minimal impact on performance.
\end{mdframed}

\begin{table*}
\centering
\caption{Objective evaluation results on MCMD. The deeper color means better performance.}
\label{tab:MCMD}
\resizebox{\textwidth}{!}{%
\begin{tabular}{cABCDABCDABCDABCDABCDABCD}
\toprule
\multirow{2}{*}{Model} & \multicolumn{4}{c}{Java} & \multicolumn{4}{c}{C\#} & \multicolumn{4}{c}{C++} & \multicolumn{4}{c}{Python} & \multicolumn{4}{c}{JavaScript} & \multicolumn{4}{c}{Average} \\ \cmidrule(r){2-5} \cmidrule(r){6-9} \cmidrule(r){10-13} \cmidrule(r){14-17} \cmidrule(r){18-21} \cmidrule(r){22-25}  
& \multicolumn{1}{c} {BLEU} & \multicolumn{1}{c} {Met.} & \multicolumn{1}{c} {Rou.} & \multicolumn{1}{c} {Cid.} & \multicolumn{1}{c} {BLEU} & \multicolumn{1}{c} {Met.} & \multicolumn{1}{c} {Rou.} & \multicolumn{1}{c} {Cid.} & \multicolumn{1}{c} {BLEU} & \multicolumn{1}{c} {Met.} & \multicolumn{1}{c} {Rou.} & \multicolumn{1}{c} {Cid.} & \multicolumn{1}{c} {BLEU} & \multicolumn{1}{c} {Met.} & \multicolumn{1}{c} {Rou.} & \multicolumn{1}{c} {Cid.} & \multicolumn{1}{c} {BLEU} & \multicolumn{1}{c} {Met.} & \multicolumn{1}{c} {Rou.} & \multicolumn{1}{c} {Cid.} & \multicolumn{1}{c} {BLEU} & \multicolumn{1}{c} {Met.} & \multicolumn{1}{c} {Rou.} & \multicolumn{1}{c} {Cid.} \\ \midrule
NNGen & 19.41 & 12.4 & 25.15 & 1.23 & 22.15 & 14.77 & 26.46 & 1.55 & 13.61 & 9.39 & 18.21 & 0.73 & 16.06 & 10.91 & 21.69 & 0.92 & 18.65 & 12.5 & 24.45 & 1.21 & 17.98 & 11.99 & 23.19 & 1.13 \\
CCT5 & 17.19 & 14.95 & 26.08 & 1.06 & 15.65 & 14.11 & 24.15 & 0.90 & 12.06 & 11.05 & 18.92 & 0.61 & 15.12 & 13.70 & 23.79 & 0.85 & 19.76 & 17.51 & 28.73 & 1.33 & 15.96 & 14.26 & 24.33 & 0.95 \\
COME & 27.17 & 23.36 & 34.59 & 1.90 & 27.29 & 23.29 & 33.33 & 1.91 & 20.8 & 17.72 & 27.01 & 1.25 & 23.17 & 19.99 & 30.48 & 1.50 & 26.91 & 23.02 & 34.44 & 1.92 & 25.07 & 21.48 & 31.97 & 1.70 \\ \midrule
GPT-3.5-Turbo & 20.24 & 17.99 & 27.8 & 1.35 & 17.81 & 15.47 & 24.38 & 1.10 & 15.29 & 13.7 & 21.49 & 0.85 & 16.5 & 15.18 & 24.13 & 0.95 & 20.38 & 18.19 & 28.26 & 1.40 & 18.04 & 16.11 & 25.21 & 1.13 \\
Claude-3-Haiku & 9.19 & 7.02 & 12.04 & 0.48 & 7.59 & 5.73 & 10.44 & 0.33 & 6.42 & 4.76 & 9.05 & 0.22 & 7.15 & 5.32 & 10.67 & 0.23 & 11.39 & 9.11 & 15.39 & 0.64 & 8.35 & 6.39 & 11.52 & 0.38 \\
Qwen1.5-7B-Chat & 6.13 & 5.38 & 10.03 & 0.19 & 5.36 & 4.81 & 8.67 & 0.14 & 5.16 & 4.56 & 7.95 & 0.13 & 5.97 & 5.58 & 10.12 & 0.14 & 7.16 & 6.78 & 11.58 & 0.23 & 5.96 & 5.42 & 9.67 & 0.17 \\ 
DeepSeek-V2-Chat & 19.36 & 17.02 & 26.5 & 1.27 & 16.78 & 14.41 & 23.19 & 1.02 & 15.48 & 13.69 & 21.76 & 0.88 & 16.53 & 15.22 & 24.11 & 0.96 & 22.08 & 19.41 & 29.52 & 1.56 & 18.05 & 15.95 & 25.02 & 1.14 \\ \midrule
CodeQwen1.5-7B-Chat & 7.28 & 5.73 & 10.96 & 0.26 & 6.63 & 5.26 & 9.91 & 0.22 & 6.00 & 4.82 & 9.31 & 0.15 & 6.16 & 4.83 & 10.45 & 0.13 & 7.79 & 6.45 & 12.27 & 0.27 & 6.77 & 5.42 & 10.58 & 0.21 \\
DeepSeek-Coder-V2-Instruct & 15.26 & 13.46 & 20.30 & 1.05 & 11.92 & 10.22 & 16.12 & 0.71 & 9.96 & 8.45 & 13.67 & 0.54 & 11.75 & 10.53 & 17.06 & 0.67 & 18.59 & 16.32 & 24.80 & 1.30 & 13.5 & 11.80 & 18.39 & 0.85 \\ \toprule
\end{tabular}%
}
\end{table*}

\begin{table*}
\centering
\caption{Objective evaluation results on MCMD-NT. The deeper color means better performance.}
\label{tab:MCMD-NT}
\resizebox{\textwidth}{!}{%
\begin{tabular}{cABCDABCDABCDABCDABCDABCD}
\toprule
\multirow{2}{*}{Model} & \multicolumn{4}{c}{Java} & \multicolumn{4}{c}{C\#} & \multicolumn{4}{c}{C++} & \multicolumn{4}{c}{Python} & \multicolumn{4}{c}{JavaScript} & \multicolumn{4}{c}{Average} \\ \cmidrule(r){2-5} \cmidrule(r){6-9} \cmidrule(r){10-13} \cmidrule(r){14-17} \cmidrule(r){18-21} \cmidrule(r){22-25}  
& \multicolumn{1}{c} {BLEU} & \multicolumn{1}{c} {Met.} & \multicolumn{1}{c} {Rou.} & \multicolumn{1}{c} {Cid.} & \multicolumn{1}{c} {BLEU} & \multicolumn{1}{c} {Met.} & \multicolumn{1}{c} {Rou.} & \multicolumn{1}{c} {Cid.} & \multicolumn{1}{c} {BLEU} & \multicolumn{1}{c} {Met.} & \multicolumn{1}{c} {Rou.} & \multicolumn{1}{c} {Cid.} & \multicolumn{1}{c} {BLEU} & \multicolumn{1}{c} {Met.} & \multicolumn{1}{c} {Rou.} & \multicolumn{1}{c} {Cid.} & \multicolumn{1}{c} {BLEU} & \multicolumn{1}{c} {Met.} & \multicolumn{1}{c} {Rou.} & \multicolumn{1}{c} {Cid.} & \multicolumn{1}{c} {BLEU} & \multicolumn{1}{c} {Met.} & \multicolumn{1}{c} {Rou.} & \multicolumn{1}{c} {Cid.} \\ \midrule
NNGen & 29.19 & 24.30 & 37 & 2.18 & 23.93 & 18.7 & 30.14 & 1.51 & 26.87 & 22.93 & 33.28 & 1.79 & 24.08 & 19.97 & 35.10 & 1.54 & 30.43 & 25.53 & 39.15 & 2.22 & 26.90 & 22.29 & 34.93 & 1.85 \\
CCT5 & 22.15 & 19.05 & 30.18 & 1.48 & 16.94 & 13.15 & 23.52 & 0.86 & 15.26 & 13.22 & 21.27 & 0.79 & 19.02 & 16.12 & 30.47 & 0.98 & 24.72 & 21.66 & 34.42 & 1.73 & 19.62 & 16.64 & 27.97 & 1.17 \\
COME & 31.46 & 26.41 & 39.53 & 2.41 & 25.60 & 20.47 & 31.68 & 1.74 & 28.83 & 25.02 & 34.90 & 1.95 & 25.95 & 22.55 & 36.78 & 1.75 & 31.30 & 27.06 & 39.77 & 2.41 & 28.63 & 24.30 & 36.53 & 2.05 \\ \midrule
GPT-3.5-Turbo & 36.31 & 34.01 & 45.61 & 2.95 & 29.67 & 26.68 & 37.67 & 2.21 & 32.22 & 30.26 & 40.51 & 2.33 & 29.91 & 28.62 & 42.39 & 2.25 & 36.83 & 34.40 & 46.82 & 3.04 & 32.99 & 30.79 & 42.60 & 2.56 \\
Claude-3-Haiku & 19.12 & 16.30 & 23.77 & 1.35 & 16.27 & 13.36 & 20.69 & 0.99 & 18.92 & 16.72 & 23.00 & 1.18 & 14.78 & 12.00 & 21.81 & 0.79 & 23.46 & 20.14 & 30.25 & 1.60 & 18.51 & 15.70 & 23.90 & 1.18 \\
Qwen1.5-7B-Chat & 9.43 & 8.22 & 14.81 & 0.41 & 7.41 & 6.33 & 11.83 & 0.23 & 8.97 & 8.89 & 14.15 & 0.40 & 7.53 & 7.21 & 12.95 & 0.23 & 8.81 & 9.16 & 15.69 & 0.29 & 8.43 & 7.96 & 13.89 & 0.31 \\
DeepSeek-V2-Chat & 36.38 & 34.53 & 45.31 & 3.00 & 29.08 & 26.15 & 36.81 & 2.11 & 32.24 & 30.71 & 40.46 & 2.37 & 30.87 & 30.02 & 42.73 & 2.36 & 37.18 & 35.03 & 47.02 & 3.09 & 33.15 & 31.29 & 42.47 & 2.59 \\ \midrule
CodeQwen1.5-7B-Chat & 8.61 & 7.10 & 13.87 & 0.31 & 8.24 & 6.14 & 11.88 & 0.27 & 12.57 & 11.00 & 16.62 & 0.60 & 7.49 & 5.81 & 11.69 & 0.24 & 9.19 & 8.30 & 14.98 & 0.34 & 9.22 & 7.67 & 13.81 & 0.35 \\
DeepSeek-Coder-V2-Instruct & 29.87 & 27.91 & 35.94 & 2.52 & 22.57 & 19.73 & 27.84 & 1.67 & 25.64 & 23.80 & 31.06 & 1.87 & 23.99 & 22.55 & 32.44 & 1.85 & 33.66 & 31.48 & 41.89 & 2.80 & 27.15 & 25.09 & 33.83 & 2.14 \\ \toprule
\end{tabular}%
}
\end{table*}

\begin{table*}
\centering
\caption{Objective evaluation results on MCMD-NL. The deeper color means better performance.}
\label{tab:MCMD-NL}
\resizebox{\textwidth}{!}{%
\begin{tabular}{cABCDABCDABCDABCDABCDABCD}
\toprule
\multirow{2}{*}{Model} & \multicolumn{4}{c}{PHP} & \multicolumn{4}{c}{R} & \multicolumn{4}{c}{TypeScript} & \multicolumn{4}{c}{Swift} & \multicolumn{4}{c}{Objective-C} & \multicolumn{4}{c}{Average} \\ \cmidrule(r){2-5} \cmidrule(r){6-9} \cmidrule(r){10-13} \cmidrule(r){14-17} \cmidrule(r){18-21} \cmidrule(r){22-25}  
& \multicolumn{1}{c} {BLEU} & \multicolumn{1}{c} {Met.} & \multicolumn{1}{c} {Rou.} & \multicolumn{1}{c} {Cid.} & \multicolumn{1}{c} {BLEU} & \multicolumn{1}{c} {Met.} & \multicolumn{1}{c} {Rou.} & \multicolumn{1}{c} {Cid.} & \multicolumn{1}{c} {BLEU} & \multicolumn{1}{c} {Met.} & \multicolumn{1}{c} {Rou.} & \multicolumn{1}{c} {Cid.} & \multicolumn{1}{c} {BLEU} & \multicolumn{1}{c} {Met.} & \multicolumn{1}{c} {Rou.} & \multicolumn{1}{c} {Cid.} & \multicolumn{1}{c} {BLEU} & \multicolumn{1}{c} {Met.} & \multicolumn{1}{c} {Rou.} & \multicolumn{1}{c} {Cid.} & \multicolumn{1}{c} {BLEU} & \multicolumn{1}{c} {Met.} & \multicolumn{1}{c} {Rou.} & \multicolumn{1}{c} {Cid.} \\ \midrule
NNGen & 27.24 & 22.82 & 32.29 & 1.90 & 25.57 & 22.05 & 29.7 & 1.65 & 26.42 & 20.88 & 35.97 & 1.64 & 24.45 & 20.07 & 30.78 & 1.63 & 29.77 & 26.05 & 32.65 & 1.96 & 26.69 & 22.37 & 32.28 & 1.76 \\
CCT5 & 31.90 & 27.31 & 37.99 & 2.20 & 33.02 & 28.92 & 37.17 & 2.19 & 32.33 & 27.92 & 43.62 & 2.24 & 29.29 & 24.58 & 37.09 & 1.98 & 28.57 & 24.62 & 31.63 & 1.68 & 31.02 & 26.67 & 37.50 & 2.06 \\
COME & 34.68 & 30.51 & 40.27 & 2.59 & 35.56 & 31.99 & 38.06 & 2.66 & 35.72 & 30.97 & 47.38 & 2.61 & 31.72 & 27.54 & 39.32 & 2.30 & 33.43 & 29.44 & 38.32 & 2.17 & 34.22 & 30.09 & 40.67 & 2.47 \\ \midrule
GPT-3.5-Turbo & 32.09 & 29.54 & 39.27 & 2.45 & 30.82 & 28.75 & 35.28 & 2.26 & 32.45 & 29.49 & 43.40 & 2.42 & 28.61 & 25.83 & 37.23 & 2.08 & 39.17 & 36.97 & 45.26 & 2.72 & 32.63 & 30.12 & 40.09 & 2.39 \\
Claude-3-Haiku & 16.66 & 14.09 & 21.02 & 1.01 & 15.71 & 13.13 & 20.86 & 0.81 & 20.11 & 16.62 & 27.16 & 1.23 & 14.81 & 12.46 & 18.47 & 0.97 & 14.69 & 12.04 & 17.52 & 0.80 & 16.40 & 13.67 & 21.01 & 0.96 \\
Qwen1.5-7B-Chat & 7.51 & 7.15 & 12.2 & 0.22 & 6.38 & 7.19 & 10.64 & 0.20 & 9.83 & 9.31 & 16.06 & 0.38 & 5.47 & 5.04 & 9.17 & 0.12 & 5.73 & 4.19 & 8.22 & 0.09 & 6.98 & 6.58 & 11.26 & 0.20 \\
DeepSeek-V2-Chat & 30.57 & 28.98 & 37.23 & 2.32 & 33.25 & 31.55 & 37.44 & 2.50 & 32.21 & 29.54 & 43.06 & 2.43 & 30.55 & 27.87 & 38.86 & 2.24 & 36.35 & 34.99 & 42.23 & 2.56 & 32.59 & 30.59 & 39.76 & 2.41 \\ \midrule
CodeQwen1.5-7B-Chat & 7.70 & 6.04 & 11.91 & 0.22 & 6.31 & 5.40 & 10.60 & 0.16 & 10.85 & 8.24 & 16.39 & 0.36 & 5.79 & 4.00 & 8.55 & 0.14 & 5.37 & 3.58 & 7.95 & 0.06 & 7.20 & 5.45 & 11.08 & 0.19 \\
DeepSeek-Coder-V2-Instruct & 27.30 & 24.86 & 32.90 & 2.05 & 26.92 & 25.10 & 30.02 & 2.03 & 27.13 & 24.33 & 35.37 & 2.03 & 24.55 & 21.56 & 30.51 & 1.79 & 31.75 & 30.16 & 36.05 & 2.24 & 27.53 & 25.2 & 32.97 & 2.03 \\ \toprule
\end{tabular}%
}
\vspace{-2ex}
\end{table*}

To explore whether the findings of LLM settings also apply to other LLMs, we conducted an additional experiment using DeepSeek-V2. Employing the same experimental setup, we replicated the experiments and found that the results closely aligned with our findings in this RQ. Detailed results are available in our artifacts \cite{LLM4CMG} due to space limitations.

\subsection{RQ2: Effectiveness of LLM}

\subsubsection{Setup}
\label{sec:RQ2_setup}
Based on the optimal settings from RQ1 (i.e., Prompt 3, 16 demonstrations, token-based selection, and sorting in ascending order), we evaluated LLMs and baselines on the test set of MCMD and MCMD-New using objective evaluation metrics.
We reused the code and the hyperparameter values released by the authors in the GitHub repositories for all baselines. Specifically, for MCMD and MCMD-NT, we reused the released model checkpoints of COME and CCT5, which were trained on MCMD. For MCMD-NL, which consists of five new languages, we trained new models for COME and CCT5, and evaluated them on the test set of MCMD-NL.

For subjective evaluation, we sampled 50 commits from each test set of MCMD and MCMD-New (10 commits per language), totaling 100 commits. We evaluated the reference messages and the messages generated by the best-performing LLMs and the best-performing baseline for these sampled commits.
To enhance the diversity of samples, we divided each test set into 50 clusters and randomly selected one commit from each cluster. We used the OpenAI embedding model \cite{embedding} for vectorization and K-means++ \cite{arthur2007k} for clustering.
In total, we sampled 400 $<$code diff, commit message$>$ pairs to score.

\subsubsection{Objective Evaluation Results}
Table~\ref{tab:MCMD}, Table~\ref{tab:MCMD-NT}, and Table~\ref{tab:MCMD-NL} show the objective evaluation results on MCMD, MCMD-NT, and MCMD-NL, respectively.

\textbf{Comparison Among LLMs.}
We find a large performance variation among different LLMs. Specifically, GPT-3.5-Turbo and DeepSeek-V2-Chat exhibit comparable performance, statistically outperforming other LLMs across all metrics and datasets. 
Given the privacy and security risks with API access to closed-source models, the open-source DeepSeek-V2-Chat model is a good choice for generating commit messages.
Interestingly, although additional pre-training on general LLMs can enhance their performance on code generation \cite{liu2024your}, it does not yield statistically significant improvements in commit message generation and may even degrade performance. 
For example, DeepSeek-V2-Chat outperforms its code-specific counterpart by 18.40\%, 21.40\%, 20.60\%, and 18.70\% in MCMD-NL in terms of BLEU, METEOR, ROUGE-L, and Cider, respectively.
This may be because code LLMs are primarily trained on code rather than commits \cite{zhu2024deepseek}, making it difficult for them to capture the subtle differences between two code snippets, which is essential to generate effective commit messages. Thus, integrating commit data into LLM training is vital to improve their performance on commit message generation.

\begin{mdframed}[nobreak=true, skipabove=5pt, skipbelow=5pt]
\textbf{Finding 4:}
GPT-3.5-Turbo and DeepSeek-V2-Chat are the best-performing LLMs for the commit message generation task. Moreover, additional code pre-training on general LLMs does not yield better performance.
\end{mdframed}

\textbf{Comparison Between LLMs and Baselines.}
We observe that the best-performing baseline, i.e., COME, surpasses the best-performing LLMs on the MCMD dataset, potentially due to the low-quality reference messages in MCMD, which will be further discussed in Sec. \ref{subjective_evaluation}.
However, on the new MCMD-NT dataset, we observe that the best-performing LLMs outperform COME on all metrics. 
The Wilcoxon signed-rank test shows that the performance difference is statistically significant at the confidence level of 95\%.
For example, GPT-3.5-Turbo outperforms COME by 15.2\%, 26.7\%, 16.6\%, and 24.9\% on average in terms of BLEU, METEOR, ROUGE-L, and Cider, respectively. This indicates that COME has poor generalization ability when faced with new data that may exhibit a different distribution from its training set.
Regarding the MCMD-NL dataset, we observe that the best-performing LLMs achieve comparable performance to COME.  It is worth noting that COME was fine-tuned on MCMD-NL, which is time-consuming and resource-intensive. In contrast, LLMs utilize only a few demonstrations to acquire task-specific knowledge without model tuning. This underscores the flexibility of LLMs in adapting to new tasks.

\begin{mdframed}[nobreak=true, skipabove=5pt, skipbelow=5pt]
\textbf{Finding 5:}
The best-performing LLMs statistically significantly outperform the best-performing baseline on MCMD-NT, indicating better generalization.
Moreover, they have comparable performance to the best-performing baseline on MCMD-NL without model tuning.
\end{mdframed}

\subsubsection{Subjective Evaluation Results} \label{subjective_evaluation}
Table~\ref{tab:subjective} shows the subjective evaluation results for the reference messages and the messages generated by the best-performing LLMs and the best-performing baseline, i.e., COME.
We observe that the best-performing LLMs outperform COME in both human and LLM-based evaluation.
The Wilcoxon signed-rank test shows that the performance difference is statistically significant at the confidence level of 95\%.
The results demonstrate the ability of LLMs to generate concise and readable commit messages with more comprehensive semantics.
Specifically, GPT-3.5-Turbo outperforms COME in terms of Informativeness, Conciseness, and Expressiveness by an average of 42.1\%/4.5\%/27.3\% and 31.8\%/7.7\%/19.7\% on the two datasets, respectively.
Notably, despite COME's superior performance in objective evaluation on MCMD, it falls behind in subjective evaluation.
This discrepancy is attributed to the low-quality reference messages in MCMD. As a learning-based method, COME tends to reproduce such low-quality messages, leading to better objective evaluation results but poorer performance in human judgment.
Additionally, we observe that the scores of reference messages in MCMD-New are higher than those in MCMD, indicating that our created MCMD-New is better in data quality.

We further investigate the correlation between automatic evaluation metrics and human evaluation by calculating Spearman's $\rho$ \cite{myers2013research} and Kendall's $\tau$ \cite{kendall1945treatment}. The results are shown in Table~\ref{tab:correlations}. 
We observe that LLM-based evaluation has much higher correlations with human judgment than other automatic evaluation metrics on both Spearman's $\rho$ and Kendall's $\tau$. This suggests that LLM-based evaluation is more reliable for evaluating the quality of commit messages.

\begin{table}
\centering
\caption{Subjective evaluation results.}
\label{tab:subjective}
\resizebox{\columnwidth}{!}{%
\begin{tabular}{cccccccc}
\toprule
\multirow{2}{*}{\makecell{Evaluation \\ by}} & \multirow{2}{*}{\makecell{Message \\ from}} & \multicolumn{2}{c}{Informativeness} & \multicolumn{2}{c}{Conciseness} & \multicolumn{2}{c}{Expressiveness} \\ \cmidrule(r){3-4} \cmidrule(r){5-6} \cmidrule(r){7-8}
 & \multicolumn{1}{c}{} & MCMD & MCMD-New & MCMD & MCMD-New & MCMD & MCMD-New \\ \midrule
\multirow{4}{*}{Human} & Reference & 3.55 & 3.65 & 3.98 & 4.25 & 4.31 & 4.49 \\
 & COME & 3.15 & 3.16 & 4.37 & 4.16 & 4.09 & 4.19 \\
 & GPT-3.5-Turbo & \textbf{4.08} & 3.99 & \textbf{4.59} & 4.51 & \textbf{4.71} & 4.61 \\
 & DeepSeek-V2-Chat & 4.05 & \textbf{4.13} & 4.44 & \textbf{4.55} & 4.69 & \textbf{4.62} \\ \midrule
\multirow{4}{*}{LLM} & Reference & 2.64 & 3.06 & 4.42 & 4.66 & 3.24 & 3.54 \\
 & COME & 2.16 & 2.30 & 4.56 & 4.52 & 2.84 & 2.86 \\
 & GPT-3.5-Turbo & \textbf{3.34} & 3.16 & \textbf{4.74} & \textbf{4.84} & 3.96 & 3.70 \\
 & DeepSeek-V2-Chat & 3.24 & \textbf{3.34} & 4.70 & 4.72 & \textbf{4.08} & \textbf{3.84} \\ \bottomrule
\end{tabular}%
}
\end{table}

\begin{mdframed}[nobreak=true, skipabove=5pt, skipbelow=5pt]
\textbf{Finding 6:}
The best-performing LLMs statistically significantly outperform the best-performing baseline in human and LLM-based evaluation.
Among automatic evaluation metrics, LLM-based evaluation has the strongest correlation with human evaluation, indicating its superior reliability in evaluating the quality of commit messages.
\end{mdframed}

\begin{table}
\centering
\caption{Spearman ($\rho$) and Kendall ($\tau$) correlations between automatic evaluation metrics and human evaluation.}
\label{tab:correlations}
\resizebox{\columnwidth}{!}{%
\begin{tabular}{ccccccccc}
\toprule
\multirow{2}{*}{Metrics} & \multicolumn{2}{c}{Informativeness} & \multicolumn{2}{c}{Conciseness} & \multicolumn{2}{c}{Expressiveness} & \multicolumn{2}{c}{Average} \\ \cmidrule(r){2-3} \cmidrule(r){4-5} \cmidrule(r){6-7} \cmidrule(r){8-9}
 & \multicolumn{1}{c}{$\rho$} & \multicolumn{1}{c}{$\tau$} & \multicolumn{1}{c}{$\rho$} & \multicolumn{1}{c}{$\tau$} & \multicolumn{1}{c}{$\rho$} & \multicolumn{1}{c}{$\tau$} & \multicolumn{1}{c}{$\rho$} & \multicolumn{1}{c}{$\tau$} \\ \midrule
BLEU & 0.2860 & 0.2058 & 0.2418 & 0.1807 & 0.2231 & 0.1674 & 0.2503 & 0.1846 \\
METEOR & 0.3213 & 0.2364 & 0.2627 & 0.1994 & 0.2111 & 0.1606 & 0.2650 & 0.1988 \\
ROUGE & 0.2745 & 0.2034 & 0.2193 & 0.1705 & 0.1907 & 0.1458 & 0.2281 & 0.1733 \\
Cider & 0.2831 & 0.2084 & 0.2334 & 0.1795 & 0.2067 & 0.1568 & 0.2410 & 0.1816 \\
GPT-4 & \textbf{0.5940} & \textbf{0.4870} & \textbf{0.3197} & \textbf{0.2822} & \textbf{0.4760} & \textbf{0.4025} & \textbf{0.4632} & \textbf{0.3906} \\ \bottomrule
\end{tabular}%
}
\vspace{-2ex}
\end{table}

\begin{figure}
    \centering
    \includegraphics[width=0.95\linewidth]{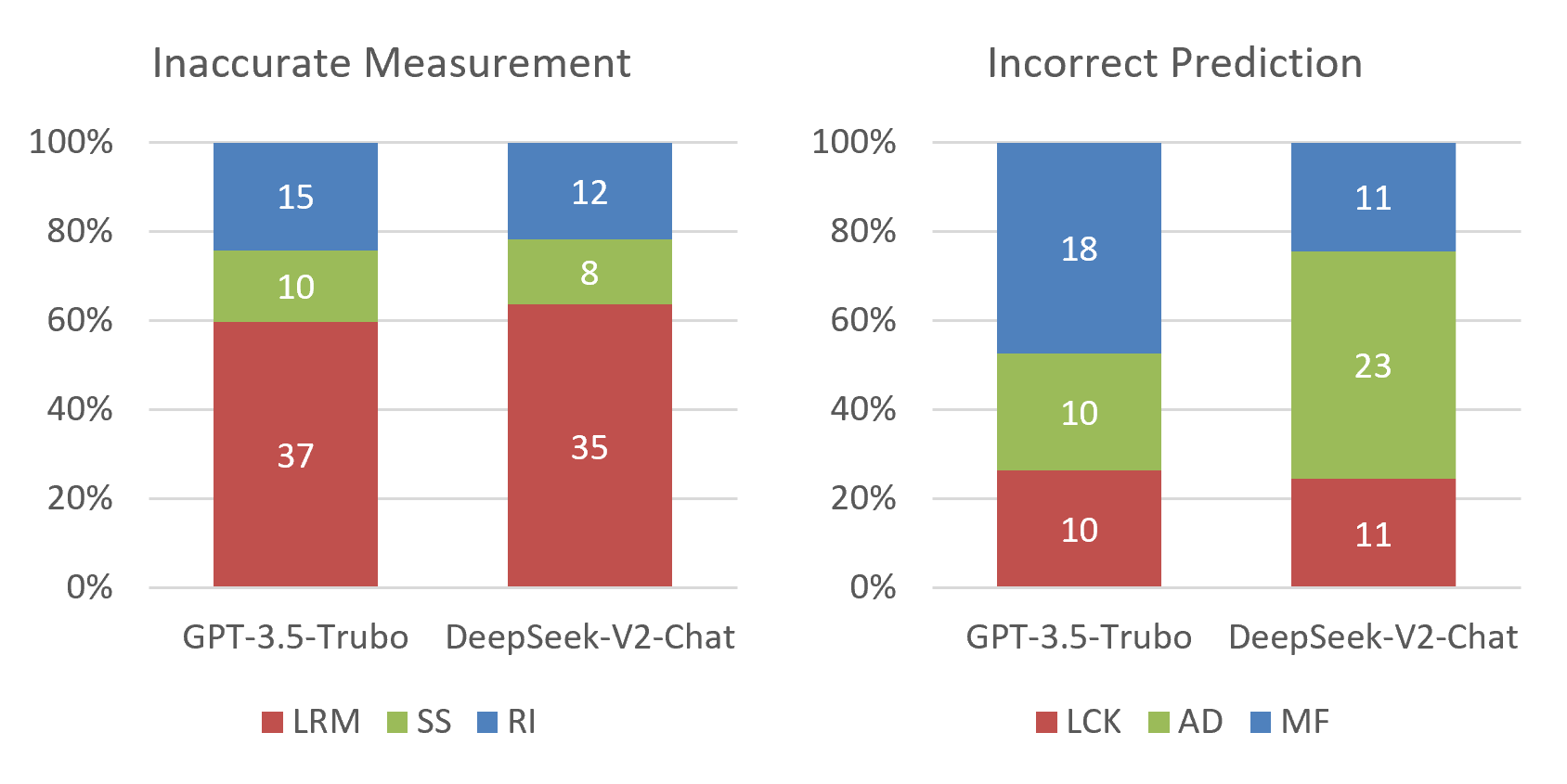}
    \caption{Results of root cause analysis.}
    \label{fig:root_cause}
\end{figure}

\subsection{RQ3: Root Cause Analysis}
\label{sec:cause}

\subsubsection{Setup}
In this RQ, we aim to further understand the root causes of underperformance for the best-performing LLMs, focusing on cases that received zero scores on all objective evaluation metrics in RQ2. We sampled 100 such cases from each of the best-performing LLMs using the same sampling method as in subjective evaluation, totaling 200 cases.
Two authors carefully read the code changes, reference messages, and generated messages by LLMs of selected cases to identify the root causes of underperformance. Following the initial analysis, the two authors discussed their disagreements to reach a consensus. When there is no agreement between the two authors, another author is introduced to discuss and resolve the disagreement.

\subsubsection{Results}
\figurename~\ref{fig:root_cause} presents the results of root cause analysis, which includes two major categories.

\begin{figure}
    \centering
    \includegraphics[width=0.95\linewidth]{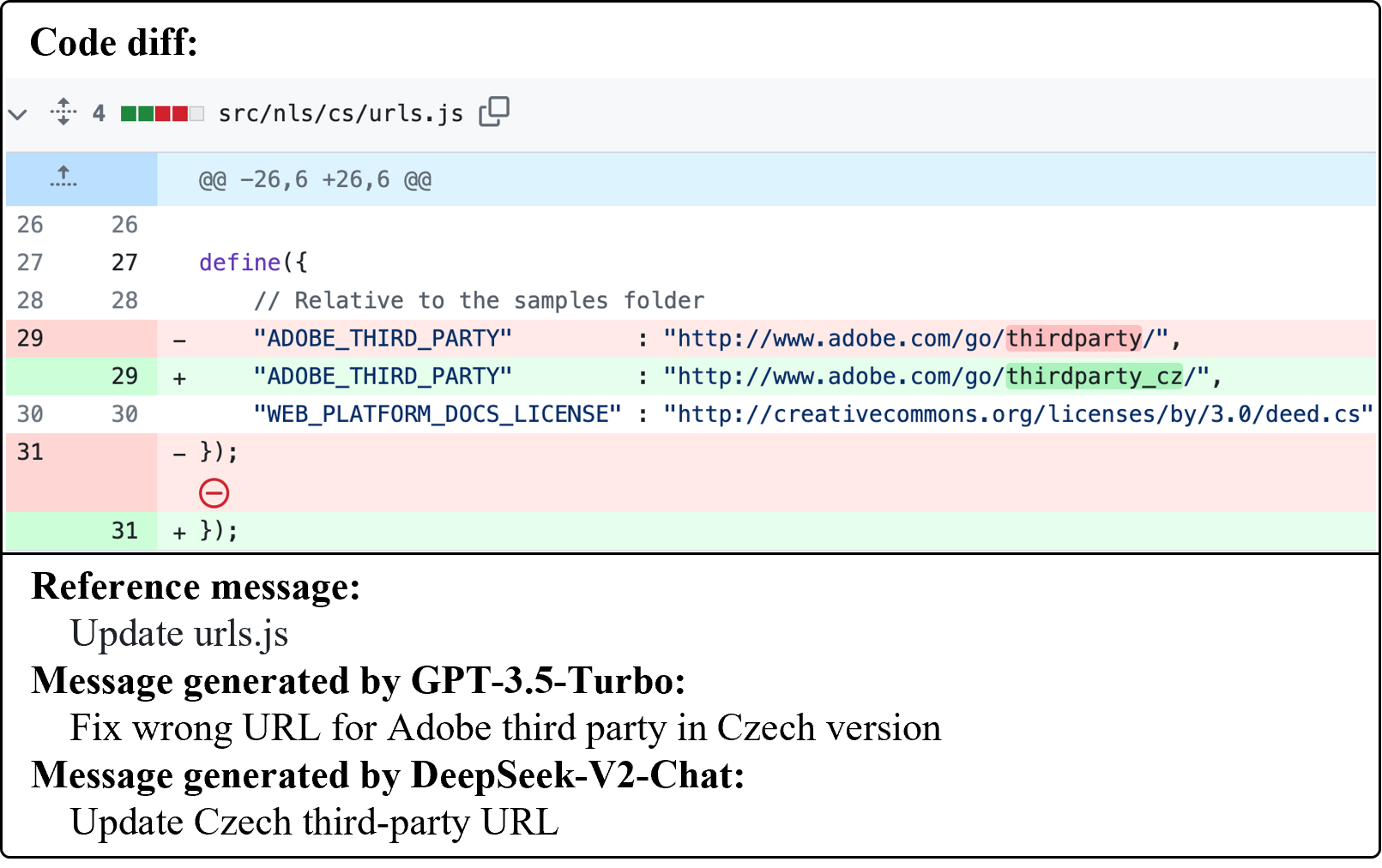}
    \caption{An example of a low-quality reference message.}
    \label{fig:case1}
    \vspace{-2ex}
\end{figure}

\begin{figure}
    \centering
    \includegraphics[width=0.95\linewidth]{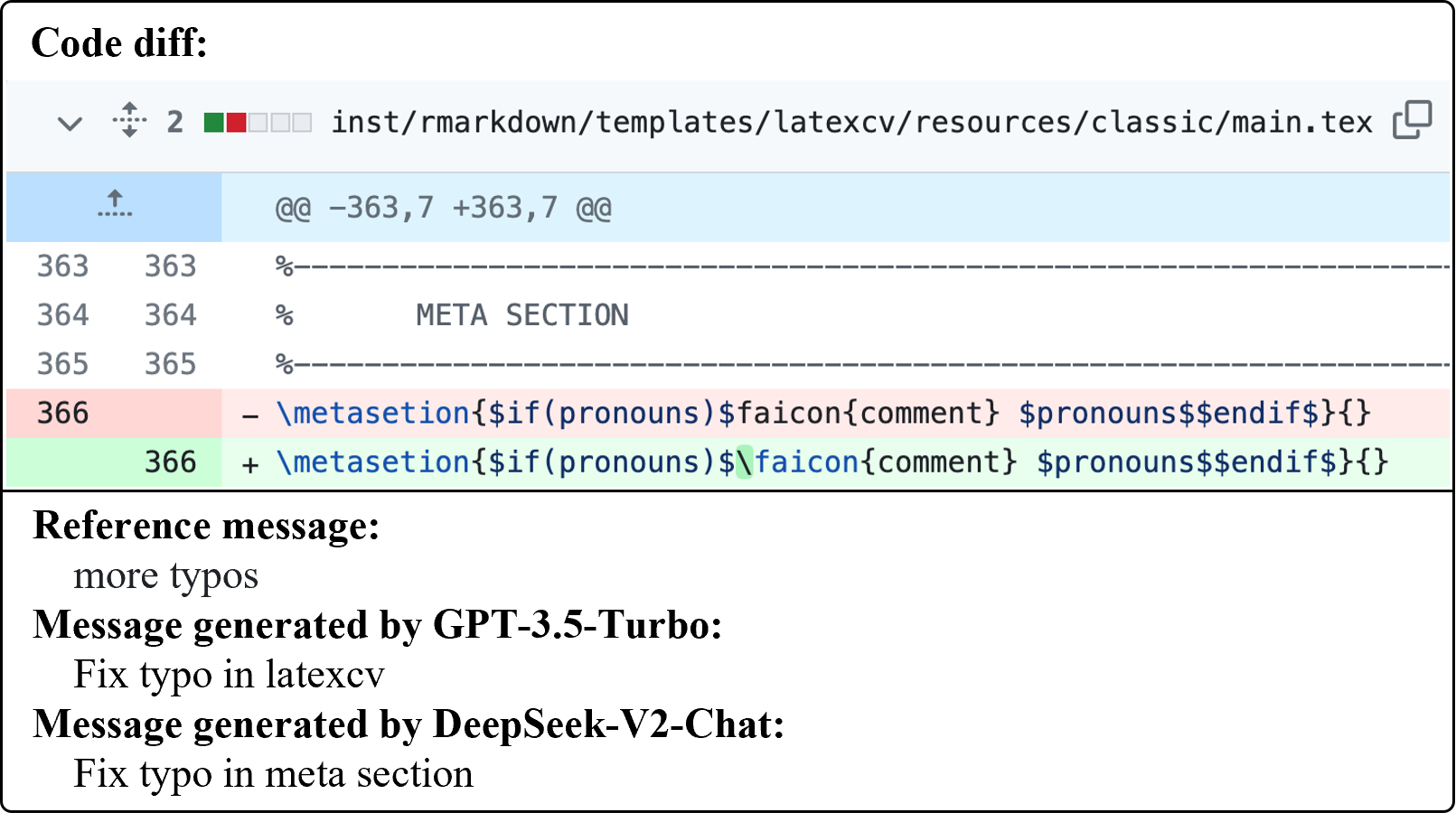}
    \caption{An example of a semantically similar message.}
    \label{fig:case2}
    \vspace{-2ex}
\end{figure}

\begin{figure}
    \centering
    \includegraphics[width=0.95\linewidth]{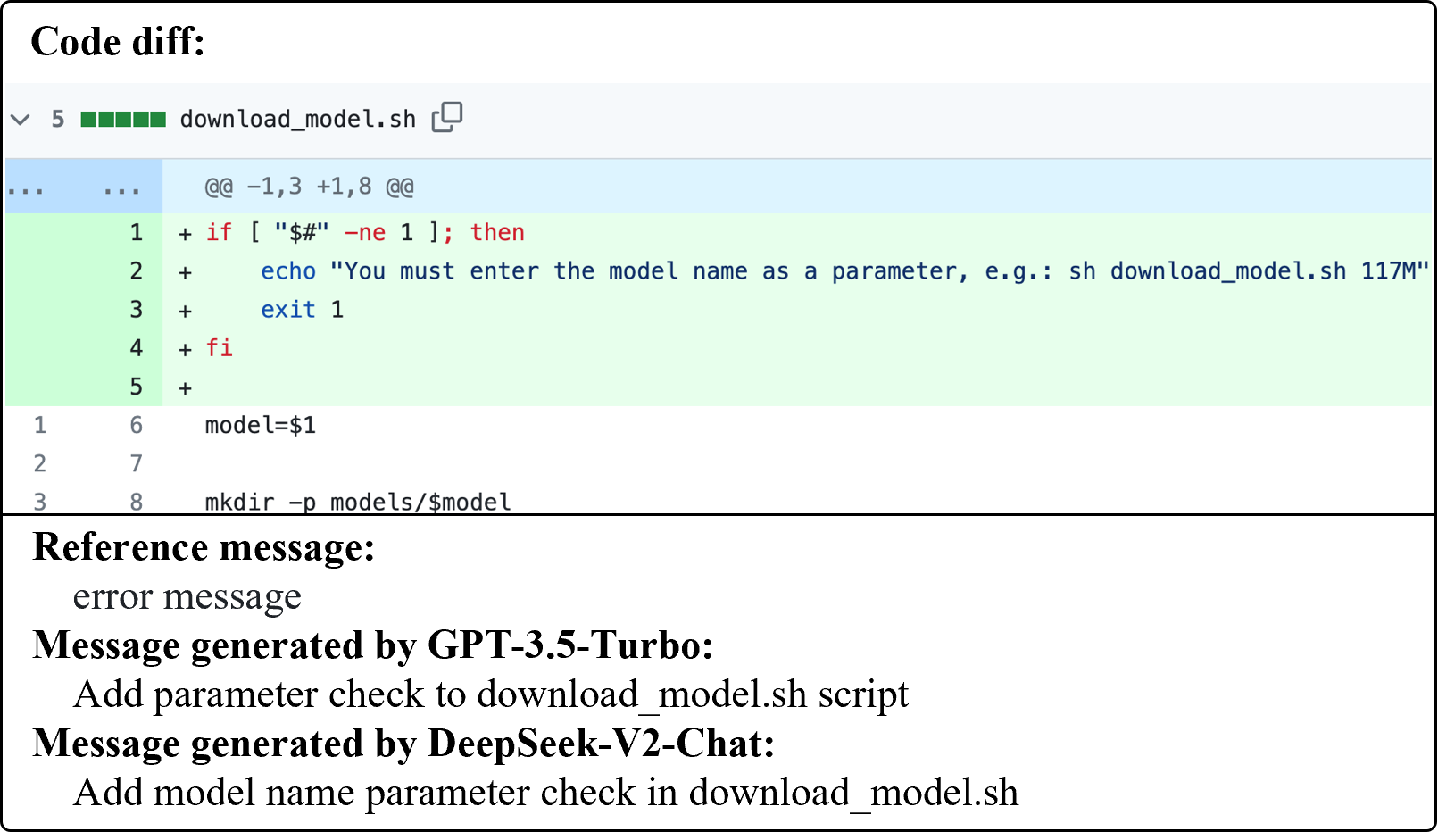}
    \caption{An example of a reasonable improvement.}
    \label{fig:case3}
    \vspace{-2ex}
\end{figure}

\textbf{Inaccurate Measurement} refers to false positives where the messages generated by LLMs are correct based on our manual inspection, but the objective evaluation metrics are low. Three types of root causes were identified in this category: 
1) \emph{Low-quality Reference Messages (LRM)}, where the reference messages have little useful information. For example, as shown in \figurename~\ref{fig:case1}, the reference message fails to describe what was changed, whereas the message generated by LLMs is more informative and gives a better summary.
2) \emph{Semantically Similar (SS)}, where the message generated by LLMs is semantically similar to the reference message. For example, as shown in \figurename~\ref{fig:case2}, all messages correctly convey the intent of the change. However, the message generated by LLMs scores zero because it fails to use the exact words present in the reference message.
3) \emph{Reasonable Improvement (RI)}, where the message generated by LLMs has a reasonable improvement over the reference message. For example, as shown in \figurename~\ref{fig:case3}, the message generated by LLMs more precisely conveys the specific intent of the code change compared to the reference message.

\begin{mdframed}[nobreak=true, skipabove=5pt, skipbelow=5pt]
\textbf{Finding 7:}
58.5\% of LLM's underperforming cases were caused by \emph{Inaccurate Measurement}, which indicates the limitation of traditional metrics and the urgent need for new metrics to accurately evaluate the performance of LLM-based commit message generation approaches.
\end{mdframed}

\textbf{Incorrect Prediction} refers to true positives where the messages generated by LLMs fail to accurately reflect the change intention. We identified three types of root causes in this category: 
1) \emph{Lack of Contextual Knowledge (LCK)} refers to cases where only code changes are not enough to provide the necessary contextual knowledge to generate correct commit messages. Additional information, such as related issues and pull requests, is required to fully understand the code changes.
2) \emph{Adverse Demonstrations (AD)} refers to cases where the reference messages in demonstrations are of low quality or the code changes in demonstrations have low similarity to the code change query. The former leads LLMs to learn and generate uninformative messages, while the latter prevents LLMs from understanding the desired behavior. Specifically, we observe that DeepSeek-V2-Chat is more susceptible to adverse demonstrations. 
3) \emph{Model Fallacy (MF)} refers to cases where LLMs fail to correctly generate commit messages due to the deficient ability of the model itself.

We further investigated potential mitigation strategies to enhance LLM's performance in cases of \emph{AD} and \emph{MF}, as \emph{LCK} requires more information.
For \emph{AD}, we refrained from providing demonstrations for LLM and prompted it to generate commit messages based solely on the code changes (i.e., zero-shot learning). The results showed that all cases were resolved.
For \emph{MF}, we utilized GPT-4, a more advanced LLM, to generate commit messages, leading to resolving 23/29 (79.3\%) cases.

\begin{mdframed}[nobreak=true, skipabove=5pt, skipbelow=5pt]
\textbf{Finding 8:}
The main root causes of LLM's underperformance are lack of contextual knowledge, adverse demonstrations, and model fallacy. Two potential mitigation strategies were providing high-quality demonstrations and improving large language models.
\end{mdframed}

\section{Discussion}

\subsection{Implications}
\textbf{Large language models are few-shot committers.}
Our empirical study reveals that LLMs are capable of generating high-quality commit messages leveraging only a few demonstrations. Our results show that the best-performing LLMs statistically significantly outperform the best-performing baseline in both human and LLM-based evaluation. This indicates that developers could utilize LLMs to help them automatically generate commit messages. Specifically, we recommend developers use the open-source DeepSeek-V2-Chat model to avoid privacy and security concerns. 
For researchers, this also indicates that the comparison with LLMs is necessary when evaluating new commit message generation approaches.
However, we also observe that additional code pre-training on general LLMs does not yield better performance on the commit message generation task. Further research is needed to explore LLMs specifically designed for code changes.

\textbf{High-quality datasets are urgently needed.}
The quality of datasets for commit message generation has not been thoroughly verified. These datasets are usually crawled from open-source projects and subjected to simple data cleaning. Our results show that 36\% of LLM’s underperformance was caused by low-quality reference messages.
Recent work \cite{tian2022makes, li2023commit} has developed automated classifiers to identify low-quality commit messages, which pave the way to construct high-quality datasets. 
Another way to enhance dataset quality is leveraging LLMs to generate high-quality pseudo-training examples based on their rich knowledge \cite{gao2023self}.
Commit messages can be written in diverse ways, but current datasets only have one reference message, which limits the diversity of the generated messages. Furthermore, our results show that 10.5\% of LLM’s underperformance was caused by lack of context.
Researchers can utilize additional information, such as issues, to construct richer datasets with multiple reference messages.

\textbf{LLM-based evaluators are promising.}
Automatic evaluation metrics, such as BLEU, simply assume that the reference messages are the gold standard and provide a quick assessment by quantifying the overlap of words or characters between the generated and the reference messages.
However, these metrics often fail to capture semantic quality like informativeness or usefulness, and their reliability can be further undermined if the reference messages are of poor quality.
Human evaluation, on the other hand, can overcome the limitation but is impractical to evaluate the whole test set due to its time-consuming and laborious nature.
Recent studies reveal that the LLM-based evaluators achieve high alignment with human judgment \cite{liu2023g, zheng2024judging}, which is in line with our finding 6.
Therefore, we advocate using LLM-based evaluators as a reliable alternative for evaluating the quality of generated messages, combining the benefits of both automatic metrics and human evaluation.
 
\subsection{Threats to Validity}
\subsubsection{Potential Data Leakage}
Since LLMs are trained on extensive data, one potential threat is data leakage, i.e., the studied LLMs have seen the commit messages in the test set. For example, the training corpus of ChatGPT includes open-source projects before Sep. 2021.
However, we observe that LLMs do not perform well with zero-shot learning, indicating a low probability of direct memorization.
To further mitigate this threat, we created a new dataset MCMD-New with open-source projects after Jan. 2022 to evaluate the studied LLMs.

\subsubsection{Randomness of LLMs and Sample Selection}
The randomness of LLM inference is a potential threat. To mitigate this, we set the temperature to 0 to generate deterministic outputs and ran experiments five times for all open-source models, which yielded more reliable and stable results. Due to budget constraints, we did not run multiple times for closed-source models in RQ2.
Moreover, the randomness of sample selection in subjective evaluation and root cause analysis is another potential threat. To mitigate the threat and ensure diverse and representative samples, we divided the dataset into multiple clusters and selected samples from each cluster.

\subsubsection{Subjectivity of Human}
The subjective nature of human evaluation in RQ2 inevitably presents a potential threat.
To reduce the threat, we invite three volunteers, including postgraduates majoring in computer science and industry professionals with relevant experience, to evaluate the commit messages, and the final result is the average of the three volunteers.
Moreover, the subjectivity of manual root cause analysis in RQ3 is another potential threat. 
To address this, we obeyed a rigorous annotation process with two authors independently annotating each sample and a third author resolving any inconsistencies or conflicts through discussion. 

\section{Related Work}

\subsection{Commit Message Generation}
Several approaches have been proposed to automatically generate commit messages, which can be categorized as rule-based, retrieval-based, and learning-based approaches.
Rule-based approaches extract information from code changes and generate commit messages with predefined templates \cite{buse2010automatically, cortes2014automatically, linares2015changescribe, shen2016automatic}. For example, Cortés-Coy et al. \cite{cortes2014automatically} extracted the stereotype, type, and impact set of a commit, then filled predefined templates with extracted information to generate commit messages.
Retrieval-based approaches find the most similar code changes in the dataset and reuse their commit messages \cite{huang2017mining, liu2018neural, huang2020learning}. For example, Liu et al. \cite{liu2018neural} represented code change as bag-of-words vectors and then calculated the cosine similarity of them to find similar code changes.
Learning-based approaches design neural machine translation models to translate code changes into commit messages \cite{liu2020atom, nie2021coregen, wang2021context, dong2022fira, jung2021commitbert, shi2022race, he2023come, tao2024kadel, lin2023cct5, loyola2017neural, jiang2017automatically, xu2019commit, liu2019generating}.
For example,
Nie et al. \cite{nie2021coregen} pre-trained the transformer model to learn the contextualized representations of code changes, and then fine-tuned the model for downstream commit message generation.
Dong et al. \cite{dong2022fira} represented code changes with a fine-grained abstract syntax tree, and used graph neural networks to extract features and generate commit messages.

\subsection{LLMs for Software Engineering}
Large language models (LLMs) have garnered significant attention and adoption in both academic and industrial domains \cite{zhao2023survey}, including software engineering \cite{hou2023large, fan2023large}, due to their exceptional performance across a wide range of applications. 
For example, Geng et al. \cite{geng2024large} investigated the feasibility of utilizing LLMs to address multi-intent comment generation.
The most related work is Gao et al. \cite{gao2023makes}, which evaluated the capability of ICL using ChatGPT on code-related tasks such as bug fixing. However, code-related tasks naturally differ from code-change-related tasks like commit message generation.
Recently, several studies have evaluated LLM-based commit message generation \cite{eliseeva2023commit, tao2024kadel, zhang2024using, lopes2024commit, zhang2024automatic, wu2024commit}.
However, these studies primarily evaluated the performance of ChatGPT in a simple zero-shot setting with basic prompts. In contrast, our study selected six mainstream LLMs and explored their performance in complex few-shot settings with different prompt and demonstration designs. In addition, we conducted subjective evaluations and analyzed the root causes of underperformance.

\section{Conclusion and Future Work}
In this paper, we conduct an empirical study on commit message generation using large language models (LLMs) via in-context learning (ICL). Specifically, we assess the impact of prompt and demonstration settings and examine LLM's effectiveness on a popular dataset and a new dataset we created. 
Our study highlights the capability of LLMs to generate commit messages through ICL and identifies several directions for future research.
In the future, we intend to develop an LLM-integrated tool that seamlessly fits into the existing software development lifecycle, aiding developers in efficiently generating high-quality commit messages.

\section*{Acknowledgment}
We thank the anonymous reviewers for their valuable comments. 
This work was supported by Ant Group Research Fund.

\clearpage
\bibliographystyle{IEEEtran}
\bibliography{mybibfile}

\end{document}